\documentclass[12pt]{article}
\usepackage{amsmath}
\usepackage{amssymb}

\title{Quantum mechanics as an approximation of statistical mechanics for classical fields}
\author{Andrei Khrennikov\\
International Center for Mathematical\\
Modeling in Physics and Cognitive Sciences,\\
MSI, University of V\"axj\"o, S-35195, Sweden\\
Email:Andrei.Khrennikov@msi.vxu.se}

\date{}
\begin{document}

\maketitle

\begin{abstract} We show that, in spite of a rather common opinion, quantum mechanics can be represented
as an approximation of classical statistical mechanics. The
approximation under consideration is based on the ordinary Taylor
expansion of physical  variables. The quantum contribution is given
by the term of the second order. To escape technical difficulties
related to the infinite dimension of phase space for quantum
mechanics, we start with a detailed presentation of our approach for
the finite dimensional quantum mechanics. We also separate real and
complex cases, because the reproduction of the complex structure of
quantum mechanics is a special problem which is not related to
approximation of classical averages. In our approach quantum
mechanics is an approximative theory. It predicts statistical
averages only with some precision. In principle, there might be
found deviations of averages calculated within the quantum formalism
from experimental averages (which are supposed to be equal to
classical averages given by our model).
\end{abstract}

\medskip


\section{Introduction}

The problem of coupling the quantum probabilistic model (which is
based on the Hilbert space calculus) and the classical probabilistic
model (which is based on the measure-theoretic calculus) was intensively discussed
already by fathers of quantum mechanics, see, e.g., the
correspondence between Einstein and Schr\"odinger \cite{ES}. This is
the problem of huge complexity, see, e.g., \cite{PAM}--\cite{KH} for
different viewpoints and debates.

Now days there is a rather common opinion that the probabilistic
structure of quantum mechanics cannot be considered simply as a special mathematical
representation of classical (measure-theoretic) probability theory. Such an opinion is based
merely on a number of no-go theorems, see appendix 2 for details. On
the other hand, there are known various prequantum models that
reproduce (at least some) features of quantum mechanics: De
Broglie's double solution theory, Bohmian mechanics, Nelson's
stochastic mechanics, t' Hooft's deterministic discrete models, see
\cite{11} --\cite{6}. All such models are either nonlocal (as
Bohmian mechanics and Nelson's stochastic mechanics) or reproduce
only some (not all) predictions of quantum mechanics. In any event
they do not contradict no-go theorems.

In author's papers \cite{KH1}, \cite{KH2} there was proposed a new prequantum
model: {\it Prequantum Classical Statistical Field Theory}, PCSFT.
This model is very natural, because this is nothing else than
standard classical statistical mechanics: a) state space is phase
space; b) variables are functions on phase space; c) statistical
states (describing ensembles of systems) are probability measures.
One important point is that phase space is {\it
infinite-dimensional.} Prequantum states (``hidden variables'') can
be represented as vector fields, $\psi(x)=(q(x), p(x)).$ However,
this is merely a technical mathematical feature of PCSFT. The crucial point
is a tricky way in which PCSFT is projected onto quantum mechanics.
This projection is {\it asymptotic.} Quantum mechanics can be
considered as an approximative theory: the quantum average given by
the von Neumann trace formula appears as the first nontrivial
contribution into the classical average (given by the Lebesgue
integral).

To clarify the main distinguishing features of our theory, PCSFT, we
shall divide its presentation into a few steps. First we consider
the finite dimensional case. Here all computations are reduced to
asymptotic expansions of simple Gaussian integrals. Then we shall
proceed to the infinite-dimensional case where we should consider
Gaussian integrals  over functional spaces.  We also
start with consideration of a toy-model of quantum mechanics over
reals, see also \cite{KH1}, \cite{KH2}. This will  simplify the
understanding of PCSFT, because the prequantum phase-space model
inducing the quantum model over complex numbers has a rather
nontrivial geometric structure. And its presentation should be
separated from asymptotic expansion of Gaussian integrals.

We remark that construction of a prequantum classical statistical
model for finite-dimensional quantum mechanics is interesting not
only from purely mathematical viewpoint. Since electron-spin and
photon-polarization can be described by two dimensional complex
spaces, in the finite-dimensional case our approach shows that that
its is possible to construct a pure classical phase-space models for
these quantum phenomena, see appendix 1.

In our approach quantum mechanics is an approximative theory. It
predicts statistical averages only with some precision. In
principle, there might be found deviations of averages calculated
within the quantum formalism from experimental averages (which are
supposed to be equal to classical averages given by our model). But
at the moment our predictions is not of a high value for
experimentalists, because PCSFT does not predict the magnitude of a
small parameter $\alpha$ in the asymptotic representation of
classical averages. In  \cite{KH1}, \cite{KH2} we speculated that 
the small parameter of PCSFT $\alpha$ (the dispersion of prequantum fluctuations) can
be chosen equal to the Planck constant $\hbar.$ But  that  speculation
was not justified. We notice that our asymptotic considerations are totally 
different from the standard considerations on the classical limit of quantum mechanics:
obtaining classical phase space mechanics as the limit of quantum mechanics when the Planck 
constant $\hbar$ (which is considered as a small parameter) goes to zero.
In our approach when the small parameter $\alpha$ goes to zero we obtain quantum 
theory as the limit case of classical (and not vice versa). In particular, neglecting by 
$\hbar$ induces neglecting by ``quantum rotation'', spin. But in PCSFT such degrees
of freedom are not neglected, see appendix 1 for details.

\section{ The Taylor approximation of averages for functions of random
variables}

Here we follow chapter 11 of the book \cite{EV} of Elena Ventzel.
This book was written in the form
of precise instructions which student should follow to solve a
problem:

``In practice we have very often situations in that, although
investigated function of random arguments is not strictly linear,
but it differs practically so negligibly from a linear function that
it can be approximately considered as linear. This is a consequence
of the fact that in many problems fluctuations of random variables
play the role of small deviations from the basic law. Since such
deviations are relatively small, functions which are not linear in
the whole range of variation of their arguments are {\it almost
linear} in a restricted range of their random changes,'' \cite{EV},
p. 238.

Let $y=f(x).$ Here in general $f$ is not linear, but it does not
differ so much from linear on some interval $[m_x-\delta,
m_x+\delta],$ where $x=x(\omega)$ is a random variable and
$$
m_x\equiv E \; x= \int x(\omega) \; d {\bf P}(\omega)
$$
is its average. Here $\delta > 0$ is sufficiently small. Student of
a military college  should approximate $f$ by using the first order
Taylor expansion at the point $m_x:$
\begin{equation}
\label{M} y(\omega) \approx f(m_x) + f^\prime (m_x) (x(\omega)-
m_x).
\end{equation}
By taking the average of both sides he obtains: \begin{equation}
\label{M1} m_y \approx f(m_x).
\end{equation}
The crucial point is that the linear term $f^\prime (m_x)
(x(\omega)- m_x)$ does not give any contribution! Further Elena
Ventzel pointed out \cite{EV}, p. 245: ``For some problems the above
linearization procedure may be unjustified, because the method of
linearization may be not produce a sufficiently good approximation.
In such cases to test the applicability of the linearization method
and to improve results there can be applied the method which is
based on preserving not only the linear term in the expansion of
function, but also some terms of higher orders.''

Let $y=f(x).$ Student now should preserve the first three terms in
the expansion of $f$ into the Taylor series at the point $m_x:$
\begin{equation}
\label{M2} y(\omega) \approx f(m_x) + f^\prime (m_x) (x(\omega)-
m_x) + \frac{1}{2} f^{\prime \prime} (m_x) (x(\omega)- m_x)^2.
\end{equation}
Hence
\begin{equation}
\label{M3} m_y \approx f(m_x) + \frac{\sigma_x^2}{2} f^{\prime
\prime} (m_x),
\end{equation}
where
$$
\sigma_x^2 = E \; (x - m_x)^2= \int \; (x(\omega) - m_x)^2 \; d {\bf
P}(\omega)
$$
is the dispersion of the random variable $x.$

Let us now consider the special case of symmetric fluctuations:
$$
m_x=0
$$
and let us restrict considerations to functions $f$ such that
$$
f(0)=0.
$$
Then we obtain the following special form of (\ref{M3}):
\begin{equation}
\label{M3H} m_y \approx  \frac{\sigma_x^2}{2} f^{\prime \prime} (0).
\end{equation}
We emphasize again that the first derivative does not give any
contribution into the average.

Thus at the some level of approximation we can calculate averages
not by using the Lebesgue integral (as we do in classical
probability theory), but by finding the second derivative. Such a
``calculus of probability'' would match well  with experiment. I
hope that reader has already found analogy with the quantum calculus
of probabilities. But for a better expression of this analogy we
shall consider the multi-dimensional case. Let now $
x=(x_1,...,x_n), $ so we consider a system of $n$ random variables.
We consider the vector average: $ m_x= (m_{x_1}, ..., m_{x_n}) $ and
the covariance matrix:
$$
B_x=(B_x^{ij}),\; B_x^{ij}= E \; (x_i -m_{x_i})\;  (x_j -m_{x_j}).
$$
We now consider the random variable
$$
y(\omega)= f(x_1(\omega), ...,x_n(\omega)).
$$
By using the Taylor
expansion we would like to obtain an algorithm for approximation of
the average $m_y.$ We start directly from the second order Taylor
expansion:
$$
y(\omega) \approx f(m_{x_1},...,m_{x_n}) + \sum_{i=1}^n
\frac{\partial f}{\partial x_i}(m_{x_1},...,m_{x_n}) (x_i(\omega)-
m_{x_i})
$$
\begin{equation}
\label{Z0} + \frac{1}{2}\sum_{i,j=1}^n \frac{\partial^2 f}{\partial
x_i\partial x_j}(m_{x_1},...,m_{x_n}) (x_i(\omega)-
m_{x_i})(x_j(\omega)- m_{x_j}),
\end{equation}
and hence:
\begin{equation}
\label{Z0A} m_y \approx f(m_{x_1},...,m_{x_1})+
\frac{1}{2}\sum_{i,j=1}^n \frac{\partial^2 f}{\partial x_i\partial
x_j}(m_{x_1},...,m_{x_1}) B_x^{ij}.
\end{equation}
By using the vector notations we can rewrite the previous formulas
as:
\begin{equation}
\label{Z1} y(\omega) \approx f(m_x) + (f^\prime(m_x), x(\omega)-
m_x) + \frac{1}{2} (f^{\prime \prime} (m_x) (x(\omega)- m_x),
x(\omega)- m_x).
\end{equation}
and
\begin{equation}
\label{Z1A} m_y \approx f(m_x) + \frac{1}{2} \rm{Tr} \; B_x
f^{\prime \prime} (m_x) .
\end{equation}
Let us again consider the special case: $m_x=0$ and $f(0)=0.$ We
have: \begin{equation} \label{Z1B} m_y \approx  \frac{1}{2} \rm{Tr}
\; B_x f^{\prime \prime} (0) .
\end{equation}
We now remark that the Hessian $f^{\prime \prime} (0)$ is {\it
always a symmetric operator.} Let us now represent $f$ by its second
derivative at zero:
$$
f \to A= f^{\prime \prime} (0).
$$
Then we see that, at some level of approximation, instead of
operation with Lebesgue integrals, one can use linear algebra:
\begin{equation}
\label{Z1BH} m_y \approx   \frac{1}{2} \rm{Tr} \; B_x A
\end{equation}

We now proceed in mathematically rigorous way, namely, we shall
estimate the reminder which was neglected in the approximative
formula for average. We also formalize correspondence between
classical and quantum statistical models.

\section{Classical and quantum statistical models}

\subsection{Classical statistical model}

Classical statistical mechanics on phase space $\Omega_{2n}={\bf
R}^n \times {\bf R}^n$ can be considered as a special classical
statistical model. In general a classical statistical model is
defined in the following way:

a).  {\it Physical states} $\omega$ are represented by points of
some set $\Omega$ (state space).

b). {\it Physical variables} are represented by functions $f: \Omega
\to {\bf R}$ belonging to some functional space
$V(\Omega).$\footnote{The choice of a concrete functional space
$V(\Omega)$ depends on various physical and mathematical factors.}

c). {\it Statistical states} are represented by probability measures
on $\Omega$ belonging to some class $S(\Omega).$

d). The {\it average} of a physical variable (which is represented
by a function $f \in V(\Omega))$ with respect to a statistical state
(which is represented by a probability measure  $\rho \in
S(\Omega))$ is given by
\begin{equation}
\label{AV0} < f >_\rho \equiv \int_\Omega f(\omega) d \rho(\omega) .
\end{equation}

\medskip

A {\it classical statistical model} is a pair
$$M=(S(\Omega), V(\Omega)).$$

\medskip

In classical statistical mechanics $\Omega= \Omega_{2n}$ is phase
space, $V(\Omega_{2n})=C^\infty( \Omega_{2n})$ is the space of all
smooth functions on phase space, $S(\Omega_{2n})$ is the space
$PM(\Omega_{2n})$ of all probability measures on phase space and the
average is given by the Lebesgue integral on the $\sigma$-algebra of
Borel subsets of $\Omega_{2n}.$

{\bf Remark 3.1.} {\small We emphasize that the space of variables
$V(\Omega)$ need not coincide with the space of all random variables
$RV(\Omega)$ -- measurable functions $\xi: \Omega \to {\bf R}.$ For
example, if $\Omega$ is a differentiable manifold, it is natural to
choose $V(\Omega)$ consisting of smooth functions; if $\Omega$ is an
analytic manifold, it is natural to choose $V(\Omega)$ consisting of
analytic functions and so on. The space of statistical states
$S(\Omega)$ need not coincide with the space of  all probability
measures $PM(\Omega).$ For example, for some statistical model
$S(\Omega)$ may consist of Gaussian measures.}

\subsection{Real finite-dimensional quantum mechanics}

We shall use a toy model of quantum mechanics which based on the
real space. Statistical features of the correspondence between a
prequantum classical statistical model and quantum mechanics are
more evident for this toy model. Denote the algebra of all $(m
\times m)$ real matrices by the symbol $M^{(r)}(m).$ We denote by
${\cal D}^{(r)}(m)$ the class of nonnegative symmetric trace-one
matrices $\rho\in M^{(r)}(m).$ We call them ``density operators.''
We denote by ${\cal L}_s^{(r)}(m)$ the class of all symmetric
matrices. In the quantum model (for the $m$-dimensional real space)
statistical states (describing ensembles of systems prepared for
measurement) are represented by density matrices and quantum
observables by matrices belonging ${\cal L}_s^{(r)}(m).$ The {\it
quantum average} of an observable $A\in {\cal L}_s^{(r)}(m)$ with
respect to a statistical state $\rho\in {\cal D}^{(r)}(m)$ is given
by the von Neumann trace class formula \cite{VN}:
\begin{equation}
\label{TR} <A>_\rho= \rm{Tr}\; \rho A.
\end{equation}
In the operator representation observables and density matrices are
corresponding classes of ${\bf R}$-linear operators. Denote the
quantum model by
$$
N_{\rm{quant}}^{(r)}= ({\cal D}^{(r)}(m), {\cal
L}_s^{(r)}(m)).
$$

If $m=1,$ then quantum observables are given by real numbers
(operators of multiplication by real numbers on the real line) and
there is only one statistical state $\rho=1.$ Here $<A>_\rho= \rho
A=A.$

\subsection{Complex finite-dimensional quantum mechanics}

Denote the algebra of all $(m \times m)$ complex matrices by the
symbol $M^{(c)}(m).$ We denote by ${\cal D}^{(c)}(m)$ the class of
nonnegative symmetric trace-one matrices $\rho\in M^{(c)}(m).$ We
call them ``density operators.'' We denote by ${\cal L}_s^{(c)}(m)$
the class of all symmetric matrices. In the quantum model (for the
$m$-dimensional complex space) statistical states (describing
ensembles of systems prepared for measurement) are represented by
density matrices and quantum observables by matrices belonging
${\cal L}_s^{(c)}(m).$ The {\it quantum average} is given  \cite{VN}
by (\ref{TR}). In the operator representation observables and
density matrices are corresponding classes of ${\bf C}$-linear
operators. Denote the quantum model by
$$
N_{\rm{quant}}^{(c)}= ({\cal D}^{(c)}(m), {\cal L}_s^{(c)}(m)).
$$
If $m=1,$ then quantum observables are given by real numbers
(operators of multiplication by real numbers on the complex plane)
and and there is only one statistical state $\rho=1.$ Here
$<A>_\rho= \rho A=A.$

\subsection{Complex quantum mechanics}

Denote by $H_c$ a complex (separable and infinite-dimensional)
Hilbert space. Denote the algebra of all bounded operators $A: H_c
\to H_c$ by the symbol ${\cal L}(H_c).$ The real linear subspace of
${\cal L}(H_c)$ consisting of {\it self-adjoint operators}  is
denoted by the symbol ${\cal L}_s(H_c).$ We denote by ${\cal
D}(H_c)$ the class of nonnegative  trace-one operators $\rho\in {\cal
L}_s(H_c).$ These are {\it von Neumann density operators.} In the
quantum model statistical states (describing ensembles\footnote{We
follow so called ensemble interpretation of quantum mechanics:
Einstein, Margenau, Ballentine, Balian, Nieuwenhuizen and many
others, see, e.g.,  \cite{BL}. By such an interpretation even a pure
state (normalized vector of $H_c)$ represents an ensemble. In the
orthodox Copenhagen interpretation a pure state represents the state
of an individual system, e.g., electron.} of systems prepared for
measurement) are represented by density operators and quantum
observables by operators from ${\cal L}_s(H_c).$ The {\it quantum
average} is given  \cite{VN} by (\ref{TR}). Denote the quantum model
by
$$
N_{\rm{quant}}(H_c)= ({\cal D}(H_c), {\cal L}_s(H_c)).
$$

\section{Taylor approximation of classical averages: one dimensional case}

States of systems are represented by real numbers, $ q \in Q={\bf
R}.$ Ensembles of such systems are described by probability measures
on the real line, {\it statistical states.} We consider a special
class of preparation procedures. They produce ensembles of systems
described by Gaussian probability distributions on $Q$ having the
zero mean value and dispersion
\begin{equation}
\label{T0} \sigma^2(\mu)= \alpha + O(\alpha^2),
\end{equation}
where as always $\vert O(\alpha^2) \vert \leq C \alpha^2$ for some
constant $C$ and a sufficiently small $\alpha.$ The crucial point is
that $\alpha$ is {\it a small parameter} of our model. Denote this
class of probability distributions by the symbol $S_{G}^\alpha(Q).$

For a probability $\mu \in S_{G}^\alpha(Q),$ we have:
\begin{equation} \label{T} d \mu (q) =  \frac{e^{\frac{-q^2}{2(\alpha +
O(\alpha^2))}}dq}{\sqrt{2\pi (\alpha + O(\alpha^2))}}.
\end{equation}
We recall that, for a probability with the zero mean value, its
dispersion is given by
\begin{equation}
\label{T1} \sigma^2(\mu) = \frac{1}{\sqrt{2\pi (\alpha +
O(\alpha^2))}} \int_{-\infty}^\infty q^2 e^{\frac{-q^2}{2(\alpha +
O(\alpha^2))}}dq .
\end{equation}
As was already pointed out, we consider $\alpha$ as a small
parameter. Therefore Gaussian probability distributions are very
{\it sharply concentrated around the point $q_0=0.$} By using the
terminology of functional analysis we say that $\{\mu\equiv
\mu(\alpha)\}$ is a $\delta$-family: $\lim_{\alpha \to 0}
\mu(\alpha)= \delta$ in the sense of theory of distributions.

In the approximation $\alpha=0$ all systems are located at a single
point, namely, $q_0.$ However, a finer description (in that $\alpha$
can not be neglected) provides the picture of Gaussian bells
concentrated nearby $q_0.$ We remark that in average a system cannot
go far away from $q_0.$ By using the Chebyshov inequality one obtain
for any $C> 0:$
\begin{equation}
\label{T2} \mu\{q: \vert q \vert > C\}\leq \frac{\alpha +
O(\alpha^2)´}{C^2} \to O, \alpha \to 0.
\end{equation}
But the probabilistic inequality (\ref{T2}) does not exclude the
possibility that some system could move far from $q_0$ (of course,
with a small probability).

We also introduce a class of {\it physical variables} in the
classical statistical model under consideration:

\medskip

a) $f\in C^\infty({\bf R}),$  a smooth function;

\medskip

b) $f(0)=0;$

\medskip

c) $\vert f^{(4)}(q) \vert \leq c_f e^{r_f \vert q\vert}, c_f, r_f
\geq 0.$

\medskip

Denote this functional space by the symbol ${\cal V}(Q), Q={\bf R}.$

 The restriction to the growth
of the fourth derivative will be used when we shall consider the
Taylor expansion of $f$ up two the fourth term. The exponential
growth implies integrability with respect to any Gaussian measure.

\medskip

{\bf Lemma 4.1.} {\it Let $f\in C^{n}$ (so it is $n$ times
continuously differentiable) and let its $n$th derivative has the
exponential growth. Then all derivatives of orders $n=0,..., n-1,$
also have the exponential growth (in particular, $f(q)$ grows
exponentially).}

{\bf Proof.} Under these conditions we can use the Taylor expansion
with the integral remainder:
\begin{equation}
\label{TT} f(q)= f(0)+ f^\prime(0) q +
\frac{f^{\prime\prime}(0)q^2}{2} + \frac{f^{(3)}(0)q^3}{3!} +...+
 \frac{q^n}{n!} \int_0^1 (1-\theta)^{n-1} f^{(n)}(\theta q) d \theta .
\end{equation}
Since the growth of  any polynomial can be compensated by decreasing
of the $e^{-r \vert q \vert},$  by using the exponential estimate
for the $n$th derivative we obtain:
\begin{equation}
\label{TT1} \vert f(q)\vert = C_{1} e^{r \vert q \vert} +
 C_{2} \frac{q^n}{n!} \int_0^1 (1-\theta)^{n-1} e^{r \vert q \theta \vert} d \theta
 \leq C e^{r \vert q \vert}.
\end{equation}
Here all constants depend on $f.$

\medskip

This simple exercise from the course of analysis will be useful in
our further considerations. We defined the following  classical
statistical model on the real line:

\medskip

A). {\it States} of systems are real numbers.

\medskip

B). {\it Statistical states} (ensembles of systems) are represented
by Gaussian probabilities having zero average and dispersion
$\sigma^2(\mu)= \alpha + O(\alpha^2), \; \alpha \to 0.$

\medskip

C). {\it Physical variables} are smooth functions with exponentially
growing fourth derivative which map zero into itself.

\medskip

We denote this model by $N_{\rm{class}}^\alpha= (S_{G}^\alpha(Q),
{\cal V}(Q)).$ As always in classical statistical physics, the
average of a physical variable $f \in {\cal V}(Q)$ with respect to
an ensemble of systems which is described by a probability $\mu \in
S_{G}^\alpha(Q)$ is given by the integral:
\begin{equation}
\label{TZ} <f>_\mu = \frac{1}{\sqrt{2\pi (\alpha + O(\alpha^2))}}
\int_{-\infty}^\infty f(q) e^{\frac{-q^2}{2 (\alpha + O(\alpha^2))}}
dq.
\end{equation}
Since $\alpha$ is a
parameter of the model, we can consider averages as functions of
$\alpha: <f>_\mu \equiv <f>_\mu(\alpha).$ We are interested in the
asymptotic expansion of averages when $\alpha\to 0.$ In particular,
such an asymptotic expansion will give us the possibility to
calculate averages approximately.

\medskip

{\bf Lemma 4.2.} {\it Let $f \in {\cal V}(Q)$ and let $\mu \in
S_{G}^\alpha(Q).$ Then }
\begin{equation}
\label{TZ1} <f>_\mu(\alpha) =\frac{\alpha}{2} f^{\prime\prime}(0) +
O(\alpha^2).
\end{equation}

{\bf Proof.} We start with the scaling of the state variable:
\begin{equation}
\label{TZ2} q= \sigma(\mu) x
\end{equation}
We have: \begin{equation} \label{TZ3} <f>_\mu(\alpha ) =
\frac{1}{\sqrt{2\pi}} \int_{-\infty}^\infty f(\sigma(\mu) x)
e^{\frac{-x^2}{2}} dx .
\end{equation}
We now expand $f(\sigma(\mu) x)$ by using the fourth order Taylor
formula with the integral remainder, see Lemma 4.1:
\begin{equation}
\label{TZ3F} <f>_\mu(\alpha) = \frac{\sigma^2(\mu)}{2} f^{\prime
\prime}(0)
\end{equation}
$$
 + \frac{\sigma^4(\mu)}{4! \sqrt{2\pi}}
\int_{-\infty}^\infty x^4 \Big(\int_0^1
(1-\theta)^{3}f^{(4)}(\sigma(\mu) x\theta) d \theta\Big)
e^{\frac{-x^2}{2}} dx.
$$
We recall that for a Gaussian measure with zero mean value all odd
momenta are equal to zero. This is an important point of our
considerations. This imply that the first nonzero contribution to
the classical average is given by the second derivative -- quadratic
term. Disappearance of the third order term implies the asymptotics
$O(\alpha^2).$ We now estimate the remainder to obtain this
asymptotics:
$$
\vert R(f, \mu) \vert \leq \frac{C \sigma^4(\mu)}{ 4!\sqrt{2\pi}}
\int_{-\infty}^\infty x^4 \Big(\int_0^1 (1-\theta)^{3} e^{r
\sigma(\mu) \vert x\vert \theta} d \theta \Big) e^{\frac{-x^2}{2}}
dx .
$$
Since we consider $\alpha$ as a small parameter, we can assume that
$\vert \sigma(\mu) \vert \leq 1$ in the exponential function. Thus:
$$
\vert R(f, \mu) \vert \leq C^\prime \sigma^4(\mu)
\int_{-\infty}^\infty x^4 e^{r \vert x\vert- \frac{x^2}{2}} dx.
$$
Since $\sigma^2(\mu)= \alpha+ O(\alpha^2),$ we have that $R(f, \mu)=
O(\alpha^2), \alpha \to 0.$

\medskip

We consider the dispersion $\sigma^2(\mu)$ as the {\it intensity of
fluctuations} in the ensemble of systems. We define the {\it
relative average} with respect to the intensity of fluctuations by
normalizing the average by the main term -- namely, $\alpha$ -- in
the intensity of fluctuations:
$$
\langle f \rangle_\mu= \frac{<f>_\mu}{\alpha}.
$$
Of course, $\langle f \rangle_\mu$ is also a function of the
parameter $\alpha:$
$$
\langle f \rangle_\mu(\alpha)= \frac{<f>_\mu(\alpha)}{\alpha}.
$$

\medskip

{\bf Corollary 4.1.} {\it Let $f \in {\cal V}(Q)$ and let $\mu \in
S_{G}^\alpha(Q).$ Then
\begin{equation}
\label{TZ4} \langle f \rangle_\mu =\frac{f^{\prime\prime}(0)}{2} +
O(\alpha).
\end{equation}
In particular,}
\begin{equation}
\label{TZ5} \lim_{\alpha\to 0} \langle f \rangle_\mu(\alpha)
=\frac{f^{\prime\prime}(0)}{2}.
\end{equation}

\medskip

{\bf Proposition 4.1.} {\it We have:}
\begin{equation}
\label{Y} \langle f \rangle_\mu = \frac{<f>_\mu}{\sigma^2(\mu)} +
O(\alpha).
\end{equation}

{\bf Remark 4.1} (About $1/2)$ {\small The second term in the Taylor
formula gives the factor 1/2 which looks rather bothering in our
asymptotic formula for the classical average. This factor will
disappear in the complex representation and the formula will become
nicer.}

We have shown that $\frac{f^{\prime\prime}(0)}{2}$ gives the
approximation of the (classical) relative average. The precision of
such an approximation is $\alpha.$ If  the level of development of
measurement technology is such that all contributions of the
magnitude $\alpha$ are neglected in measurements, then averages can
be calculated by using the following simple rule:
\begin{equation}
\label{TZ6} \langle f \rangle_\mu^{\rm{approx}} =
\Big[\frac{<f>_\mu}{\sigma^2(\mu)}\Big]^{\rm{approx}}
=\frac{f^{\prime\prime}(0)}{2}.
\end{equation}
At the first sight such averages have nothing to do with classical
averages given by integrals. There could be even presented an
interpretation of physics claiming that rules of classical
probability theory are violated  and relating the exotic rule
(\ref{TZ6}) for calculating of averages to special features of
systems under consideration (and not to a special approximation
procedure for averages).

Finally, we remark that calculation of averages by (\ref{TZ6}) is
essentially simpler than  classical probabilistic averages given by
Lebesgue integrals.

\section{Taylor approximation of classical averages: multidimensional case}

States are vectors $ q  \in Q={\bf R}^m;$ statistical states are
Gaussian distributions with the zero mean value and the {\it
dispersion} $\sigma^2(\mu)= \alpha + O(\alpha^2).$ Denote this class
of probabilities by the symbol $S_{G}^\alpha(Q).$ We introduce the
scalar product and norm on $Q:$
$$
(\xi, q)=\sum_{j=1}^m \xi_j q_j, \; \;\Vert q \Vert^2 = \sum_{j=1}^m q_j^2.
$$
If a Gaussian measure $\mu$ is nondegenerate (so the measure of any
open set is positive), then
$$
d\mu(q) = \frac{e^{-\frac{1}{2} (B^{-1} q, q)} \; d
q}{\sqrt{(2\pi)^m \det B}},
$$
where $B$ is a positive operator (we consider everywhere only
Gaussian measures with zero mean values). If $\mu \in
S_{G}^\alpha(Q)$ and nondegenerate, then
$$
\sigma^2(\mu)= \frac{1}{\sqrt{(2\pi)^m \det B}} \int_{{\bf R}^m}
\Vert q \Vert^2 \; e^{-\frac{1}{2} (B^{-1} q, q)} \; d q =  \alpha +
O(\alpha^2).
$$
In the general case the easiest way to define a Gaussian measure is
to use its Fourier transform:
$$
\tilde{\mu}(\xi)= \int_{{\bf R}^m} e^{i (\xi,q)}  \; d \mu(q)=
e^{-\frac{1}{2}(Bq,q)},
$$
where $B=\rm{cov}  \; \mu$ is the {\it covariance operator:}
$$
(B\xi_1, \xi_2) = \int_{{\bf R}^m} (\xi_1,q)  \; (\xi_2,q)  \; d
\mu(q).
$$
We remark that by definition  a covariance operator is {\it
positively defined and symmetric.}

\medskip

{\bf Lemma 5.1.} {\it Let $\mu$ be a Gaussian measure with the zero
mean value and let $A$ be a symmetric operator. Then
\begin{equation} \label{EQ} \int_{{\bf R}^m} (Aq,q)   \;d \mu(q) =
\rm{Tr}  \;B A,
\end{equation}
where $B= \rm{cov}  \; \mu.$}

\medskip

To prove this lemma we should just expand the quadratic form
$(Aq,q)$ with respect to an orthonormal basis.

\medskip

{\bf Corollary 5.1.} {\it We have}
\begin{equation}
\label{EQ1} \sigma^2(\mu) = \int_{{\bf R}^m} \Vert q \Vert^2   \; d
\mu(q) = \rm{Tr}  \; B.
\end{equation}

Thus, for $\mu \in S_{G}^\alpha(Q),$
$$
\rm{Tr}  \; \rm{cov}  \; \mu= \alpha + O(\alpha^2).
$$
We now define a class of physical variables -- ${\cal V}(Q):$ a)
$f\in C^\infty({\bf R}^m);$ b) $f(0)=0;$ c) $\Vert f^{(4)}(q) \Vert
\leq c_f e^{r_f \Vert q\Vert}, c_f, r_f \geq 0.$

For a function $f: {\bf R}^m \to {\bf R},$ its $n$th derivative is a
(symmetric) $n$-linear functional, $f^{(n)}(q): {\bf R}^m\times
...\times {\bf R}^m \to {\bf R}.$ The norm of this functional is
given by
$$
\Vert f^{(n)}(q) \Vert =\sup_{\Vert h_j \Vert =1} \vert
f^{(n)}(q)(h_1,...,h_n)\vert.
$$
The norm can be estimated by partial derivatives:
$$
\Vert f^{(n)}(q) \Vert \leq \max_{\alpha_1 +...+ \alpha_n=
n}\Big\vert \frac{\partial^n f(q)}{\partial q_1^{\alpha_1} ...
\partial q_n^{\alpha_n}}\Big\vert.
$$
It is easy to generalize Lemma 4.1 to the multidimensional case.

Thus we have defined the following classical statistical model:
$N_{\rm{class}}= (S_{G}^\alpha(Q), {\cal V}(Q)).$

\medskip

{\bf Lemma 5.2.} {\it Let $f \in {\cal V}(Q)$ and let $\mu \in
S_{G}^\alpha(Q).$ Then
\begin{equation}
\label{TZ1Z} <f>_\mu(\alpha)\equiv \int_{{\bf R}^m} f(q)  \; d
\mu(q) = \frac{\alpha}{2} \rm{Tr} \; \rho f^{\prime\prime}(0) +
O(\alpha^2),
\end{equation}
where $\rho$ is a density operator; in fact, $\rho= \rm{cov}\;
\mu/\alpha.$}

{\bf Proof.} By using the scaling of the state variable (\ref{TZ2})
and by expanding $f(\sigma(\mu) x)$ on the basis of the fourth order
Taylor formula with the integral remainder we obtain:
\begin{equation}
\label{TZ3Z} <f>_\mu(\alpha) = \frac{\sigma^2(\mu)}{2}  \rm{Tr}
\;\rho f^{\prime \prime}(0)
\end{equation}
$$
 + \frac{\sigma^4(\mu)}{4!}
\int_{{\bf R}^m} \Big(\int_0^1 (1-\theta)^{3}\; f^{(4)}(\sigma(\mu)
x\theta)(q,q,q,q) \;d \theta\Big) \;d \mu_{\rm{scal}}(x),
$$
where $ \mu_{\rm{scal}}$ is a normalized Gaussian measure -- the
image of $\mu$ under the scaling (\ref{TZ2}). We now estimate the
remainder:
$$
\vert R(f, \mu) \vert \leq \frac{C \sigma^4(\mu)}{ 4!} \int_{{\bf
R}^m} \Vert x\Vert^4 \;\Big(\int_0^1 (1-\theta)^{3} \;e^{r
\sigma(\mu) \Vert x\Vert \theta}\; d \theta \Big)\; d
\mu_{\rm{scal}}(x).
$$
Thus
$$
\vert R(f, \mu) \vert \leq C^\prime \sigma^4(\mu) \int_{{\bf R}^m}
\Vert x\Vert^4 \;e^{r \Vert x\Vert}\;d \mu_{\rm{scal}}(x).
$$
We have that $R(f, \mu)= O(\alpha^2), \alpha \to 0.$

\medskip

{\bf Corollary 5.2.} {\it Let $f \in {\cal V}(Q)$ and let $\mu \in
S_{G}^\alpha(Q)$ be nondegenerate. Then
\begin{equation}
\label{TZ1Z0} \frac{1}{\sqrt{(2\pi)^m \;\det B}} \int_{{\bf R}^m}
f(q)\; e^{\frac{1}{2} (B^{-1} q, q)} \;d q = \frac{\alpha}{2}
\rm{Tr}\; \rho f^{\prime\prime}(0) + O(\alpha^2).
\end{equation}
where $\rho= B/\alpha.$}

As in the one-dimensional case, we introduce the relative average:
$$
\langle f \rangle_\mu\equiv \frac{<f>_\mu}{\alpha}=\frac{\int_{{\bf
R}^m} \; f(q) \; d\mu(q)}{\int_{{\bf R}^m} \Vert q \Vert^2\;
d\mu(q)} +O(\alpha).
$$
In the case of a nondegenerate Gaussian measure we have:
$$
\langle f \rangle_\mu =\frac{ \int_{{\bf R}^m} f(q)\; e^{\frac{1}{2}
(B^{-1} q, q)} \;d q}{ \int_{{\bf R}^m} \Vert q \Vert^2 \;
e^{\frac{1}{2} (B^{-1} q, q)} \;d q} +O(\alpha).
$$

{\bf Corollary 5.3.} {\it Let $f \in {\cal V}(Q)$ and let $\mu \in
S_{G}^\alpha(Q).$ Then}
\begin{equation} \label{KKK}
\langle f \rangle_\mu=\frac{1}{2} \rm{Tr}\; \rho \;
f^{\prime\prime}(0) + O(\alpha).
\end{equation}

Thus if one neglects by terms of the magnitude $\alpha,$ it is
possible to use the following approximative calculus of averages:
\begin{equation}
\label{KK0} \langle f \rangle_\mu^{\rm{approx}} = \frac{1}{2}
\rm{Tr}\; \rho A,
\end{equation}
where $A=f^{\prime\prime}(0)$ and $\rho=\rm{cov}\; \mu_{\rm{scal}}.$
This is nothing else than the von Neumann trace formula for quantum
averages, see (\ref{TR}). To proceed more formally, we consider
maps:
\begin{equation}
\label{KK1} T: S_{G}^\alpha(Q) \to {\cal D}^{(r)}(m), \rho=T(\mu)=
\rm{cov}\; \mu_{\rm{scal}};
\end{equation}
\begin{equation}
\label{KK2} T: {\cal V}(Q)) \to {\cal L}_s^{(r)}(m)),
A=T(f)=f^{\prime\prime}(0)
\end{equation}
(we recall that Hessian is always a symmetric matrix).

\medskip

{\bf Theorem 5.1.} {\it The maps (\ref{KK1}), (\ref{KK2}) project
the classical statistical model $N_{\rm{class}}= (S_{G}^\alpha(Q),
{\cal V}(Q))$ onto the quantum model $N_{\rm{quant}}^{(r)}= ({\cal
D}^{(r)}(m), {\cal L}_s^{(r)}(m))$  in such a way that classical and
quantum averages are coupled by the asymptotic equality:}
\begin{equation}
\label{KK3} \langle f \rangle_\mu= \frac{1}{2} <T(f)>_{T(\mu)} +
O(\alpha).
\end{equation}

\section{Degenerate Gaussian measures and ``pure states''}

Consider a Gaussian measure $\mu \in S_{G}^\alpha(Q)$ which is
concentrated on a linear subspace $Q_0$ of $Q.$ So  it is
nondegenerate on $Q_0.$ Here $P: Q \to Q_0$ is the orthogonal
projector onto $Q_0.$ Denote by $B_0$ the covariance matrix of the
restriction of $\mu$ onto $Q_0.$ Then $B_0
>0$ and $B= \rm{cov}\; \mu= P B_0P.$ Let us now make the scaling (\ref{TZ2}).
Then $\rho=\rm{cov}\; \mu_{\rm{scal}}= P \rho_0 P,$ where $\rho_0=
\rm{cov}\; \mu_{\rm{scal}}\vert_{Q_0}.$

Thus for any symmetric matrix $A$ we have  $\rm{Tr}\; \rho A=
\rm{Tr}\; \rho_0 (P A P).$ Suppose now that $\rho_0=
\frac{I}{\rm{dim}\; Q_0},$ where $I:Q_0\to Q_0$ is the unit
operator. Then $\rm{Tr}\; \rho A= \frac{1}{\rm{dim}\; Q_0}\rm{Tr}\;
P A P.$ We are especially interested in measures concentrated on one
dimensional subspaces, $Q_0\equiv Q_\Psi=\{q= c\Psi, c \in {\bf
R}\},$ where $\Psi$ has the norm one. Here $P\equiv P_\Psi$ is the
one dimensional projector $P_\Psi: Q \to Q_\Psi, P_\Psi \phi= (\phi,
\Psi) \Psi,$ and hence
\begin{equation} \label{KK4}
\rm{Tr}\; \rho A= \rm{Tr}\; P_\Psi A P_\Psi = (A\Psi,
\Psi).
\end{equation}

Denote a probability $\mu \in \in S_{G}^\alpha(Q)$ which is
concentrated on the one dimensional subspace $Q_\Psi$ by the symbol
$\mu_\Psi.$ We obtained the following simple result:

{\bf Proposition 6.1.} {\it For any $\mu_\Psi,$ we have}
\begin{equation} \label{KK5}
\langle f \rangle_{\mu_\Psi}=\frac{1}{2} (f^{\prime\prime}(0) \Psi,
\Psi)  + O(\alpha).
\end{equation}

Thus approximately we have:
\begin{equation} \label{KK6}
\langle f \rangle_{\mu_\Psi}=\frac{1}{2} (A \Psi,  \Psi), \;
A=f^{\prime\prime}(0).
\end{equation}
But the right-hand side of this equality is nothing else than the
well known quantum formula for the average of the quantum observable
$A$ with respect to the pure state $\Psi,$  \cite{VN}. In our
approach this quantum formula arose as the approximation of the
classical average with respect to a Gaussian ensemble. The only
distinguishing feature of such an ensemble is that  all systems have
states proportional to the vector $\Psi$ (with the probability one).
Of course, one can consider the projective space and then all those
systems will have the same coordinate. However, real coordinates of
systems are different.

\medskip

{\bf Conclusion.} {\it Quantum averages with respect so called pure
states can be easily reproduced as approximations of ordinary
ensemble averages with respect to one dimensional Gaussian
distributions.}

\section{Prequantum phase space -- the two dimensional case}

In previous sections we considered the  prequantum toy model in that
the phase space structure was not taken into account. The
corresponding quantum model was over the reals, see also \cite{KH1}, \cite{KH2}. On the
other hand, physical reality is described by the classical phase
space mechanics and the complex  quantum mechanics. We shall see
that it is possible to create a prequantum phase space model
reproducing the complex quantum mechanics. The crucial point is that
classical variables and statistical states -- functions and measures
on phase space -- should be {\it invariant with respect to a special
group of transformations of phase space.}

This {\it fundamental prequantum group} is very simple -- the {\it
special orthogonal group} $SO(2),$ the group of rotations of phase
space.

\medskip

States of systems are now represented by points $\psi=(q,p)\in
\Omega= Q\times P,$ where $Q=P={\bf R}.$ Here the $q$ is the
position and the $p$ is momentum, so $\Omega$ denotes phase space.
Statistical states are represented by Gaussian $SO(2)$-invariant
measures having zero mean value and dispersion
\begin{equation} \label{D}
\sigma^2(\mu)= 2 \alpha+ O(\alpha^2);
\end{equation}
physical variables are by $SO(2)$-invariant maps, $f: \Omega \to
{\bf R},$ which satisfy conditions a), b), c) specifying variables
in the real case. Denote these classes of measures and functions,
respectively, $S_G^\alpha(\Omega \vert SO(2))$ and ${\cal V}(\Omega
\vert SO(2)).$

The appearance of the factor 2 has the following motivation: there
are two contributions into fluctuations -- fluctuations of positions
and momenta. We shall see that they are equally distributed.
Therefore it is natural to consider as a small parameter of the
model the dispersion  of e.g. the $q$-fluctuations (which equals to
the dispersion of the $p$-fluctuations).

We consider the classical model $N_{\rm{class}}= (S_G^\alpha(\Omega
\vert SO(2)), {\cal V}(\Omega \vert SO(2)).$ As in the real case, we
can obtain the asymptotic expansion of the classical averages, see
(\ref{KKK}). However, in quantum mechanics we consider the complex
structure. We would like to recover it in our classical model. To do
this, we shall study in more detail properties of classical
probabilities and variables.

A measure $\mu$ is invariant if for any $u\in SO(2):$
\begin{equation} \label{KK11}
\int_{{\bf R}^2} f(uq) d\mu(q)= \int_{{\bf R}^2} f(q) d\mu(q).
\end{equation}
For a Gaussian measure $\mu$ with the covariance matrix $B,$ this is
equivalent to the condition:
\begin{equation} \label{KK7G}
[u, B]=0, \; u\in SO(2).
\end{equation}

Let $f$ be a two times differentiable invariant map, so $f(u\psi)=
f(\psi),$ for any $u\in SO(2).$ By representing
\begin{equation} \label{RER}
u=u_\theta= \left(
\begin{array}{ll}
 \cos \theta & - \sin \theta\\
\sin \theta & \cos \theta
\end{array}
 \right ),
\end{equation}
we have that
\begin{equation} \label{KK9G}
f(\cos \theta q - \sin \theta p, \sin \theta q + \cos \theta p)
=f(q, p).
\end{equation}
This is a rather strong constraint determining a very special class
of maps. In particular, we obtain: $ u^* \nabla f(u \psi) =\nabla
f(\psi)$ and $u^* f^{\prime \prime}(u \psi) u = f^{\prime
\prime}(\psi).$ Hence $u^* \nabla f(0) =\nabla f(0)$ for any
rotation, and thus
\begin{equation} \label{KK15}
\nabla f(0)=0
\end{equation}
and
\begin{equation} \label{KK8G}
[f^{\prime \prime}(0), u]=0, \; u\in SO(2).
\end{equation}

It is convenient to introduce the {\it commutator} of the set
$SO(2)$ in the algebra of all two by two matrices $M^{(r)}(2):$
$$
SO^\prime(2)=\{ A \in M^{(r)}(2): [A, u]= 0, u \in SO(2)\}
$$
We remark that a generator of $SO(2)$ can be chosen as the
symplectic operator:
 \[J= \left( \begin{array}{ll}
 0&1\\
-1&0
\end{array}
 \right ).
 \]
Therefore the commutator of $SO^\prime(2)$ coincides with the
commutator of $J: \{ J\}^\prime= \{ A \in M^{(r)}(2): [A, J]= 0\}.$

\medskip

{\bf Proposition 7.1.} {\it Let $\mu \in S_G^\alpha(\Omega \vert
SO(2))$and let  $f\in {\cal V}(\Omega \vert SO(2)).$ Then
$B=\rm{cov}\; \mu$ and $A= f^{\prime \prime}(0)$ belong to
$SO^\prime(2).$}

\medskip

{\bf Lemma 7.1.} {\it A matrix $A$ belongs to the commutator
$SO^\prime(2)$ iff
\begin{equation} \label{U}
A= \left(
\begin{array}{ll}
 R&-S\\
S & R
\end{array}
 \right ).
\end{equation}
 If $A$ is also symmetric, then it is diagonal:
$A= \left(\begin{array}{ll}
 R & 0\\
 0 & R
\end{array}
 \right ).$ In particular, its trace is given by}
\begin{equation} \label{KK14}
\rm{Tr} A=2 R
\end{equation}

Thus if $\mu \in S_G^\alpha(\Omega \vert SO(2)),$ then its
covariance matrix is diagonal $B= \left(\begin{array}{ll}
 b & 0\\
 0 & b
\end{array}
 \right ),$ where $2b=\alpha +O(\alpha^2).$ Fluctuations of
the
 coordinate $q$  and the momentum $p$ are independent and equally
 distributed:
 $$
 d\mu(q)= \frac{1}{2\pi b} \exp\{-\frac{q^2+p^2}{2b}\} dq.
 $$

Denote the marginal distributions of $\mu$ by the symbols $\mu_q$
and $\mu_p,$ respectively. Then
$$
\sigma^2(\mu_q )=\frac{1}{\sqrt{2\pi b}} \int_{-\infty}^{+\infty}
q^2\; \exp\{-\frac{q^2}{2b}\} dq =
\sigma^2(\mu_p)=\frac{1}{\sqrt{2\pi b}}\int_{-\infty}^{+\infty}
p^2\; \exp\{-\frac{p^2}{2b}\} dp.
$$
Hence
$$
\sigma^2(\mu_q)= \sigma^2(\mu_p)= \frac{1}{2} \sigma^2(\mu)= \alpha
+O(\alpha^2).
$$

{\bf Proposition 7.2.} {\it Let $f\in {\cal V}(\Omega \vert SO(2)).$
Then all its odd derivatives at the point $q_0=0$ are equal to
zero.}

{\bf Proof.} For any vector $\phi\in \Omega,$  we have
$f^{(2n+1)}(u\psi)(u\phi,...,u\phi)$\\$=
f^{(2n+1)}(\psi)(\phi,...,\phi),$ for any rotation $u.$ We choose
$u=J.$ Then:
$$
(-1)^{2n+1} f^{(2n+1)}(0)(\phi,...,\phi)=
f^{(2n+1)}(0)(J^2\phi,...,J^2\phi)$$
$$
=f^{(2n+1)}(0)(J\phi,...,J\phi)= f^{(2n+1)}(0)(\phi,...,\phi).
$$
Thus $f^{(2n+1)}(0)(\phi,...,\phi)=0$ for any vector $\phi\in
\Omega.$

\medskip

For a function $f\in {\cal V}(\Omega \vert SO(2)),$ its Hessian has
the form $f^{\prime \prime}(0)= \left(\begin{array}{ll}
 R & 0\\
 0 & R
\end{array}
 \right ),$ where $R \in {\bf R},$ and hence:
$$
f(q,p)= \frac{R(q^2+ p^2)}{2} +O(\alpha^2).
$$

We remark that, in spite of the coincidence of commutators, the
$SO(2)$-invariance is not equivalent to the $J$-invariance (the
later was used as the basis of the theory in \cite{KH3}).

\medskip

{\bf Example 7.1.} Let $f(q,p)=q^3 p - q p^3= qp (q^2- p^2).$ Then
$f(J\psi)=f(\psi).$ But take $\theta=\pi/4.$ Here
$u=\frac{1}{2}\left(\begin{array}{ll}
 1 & -1\\
 1 & 1
\end{array}
 \right ).$ Hence $u \left(\begin{array}{l}
 q\\
 p
\end{array}
 \right ) =  \left(\begin{array}{l}
 (q-p)/2\\
 (q+p)/2
\end{array}
 \right ).$ Thus $f(u\psi)= (q-p)(q+p) (q-p- q-p)(q-p+q+p)/16=
 -qp ((q^2- p^2)/4.$

\medskip

We are now  completely ready to recover the complex structure of
quantum mechanics. By Lemma 7.1 any matrix belonging $SO^\prime(2)$
can be represented in the form: $A= R\; I \; + \; S\; (-J).$  By
mapping $I$ into 1 and $(-J)$ into $i$ we obtain a map of  the
commutator $SO^\prime(2)$  onto the set of complex numbers ${\bf
C}:$
\begin{equation} \label{ZX}
j: SO^\prime(2) \to {\bf C}, z=j(A)= R+iS .
\end{equation}
This is the isomorphism of two fields.

In particular, a symmetric matrix $A= \left(
\begin{array}{ll}
 R& 0\\
0 & R
\end{array}
 \right )$ is represented by the real number $j(A)=R.$
This is the operator of multiplication by $R.$ The trace of this
operator in the one dimensional complex space ${\bf C}$ (with the
scalar product, $(z,w)= z\bar{w})$ equals $R.$ By (\ref{KK14}) we
obtain that
\begin{equation} \label{ZX1}
\rm{Tr} \;A =2 \rm{Tr}\; j(A),
\end{equation}
where at the left-hand side we have the real trace and at the
right-hand side -- the complex trace. Now we can write the basic
asymptotic equality for averages in the complex form. In the funny
way the Taylor factor $\frac{1}{2}$ disappears through the
transition form the real to complex structure, see (\ref{ZX1}).

\medskip

{\bf Lemma 7.2.} {\it Let $f \in {\cal V}(\Omega \vert SO(2))$ and
let $\mu \in S_G^\alpha(\Omega \vert SO(2)).$ Then
\begin{equation}
\label{P0} <f>_\mu(\alpha)\equiv \int_{{\bf R}^2} f(q,p)  \; d
\mu(q,p) = \alpha \; j(f^{\prime\prime}(0)) + O(\alpha^2).
\end{equation}
}

{\bf Proof.} We make the scaling of the state variable:
\begin{equation}
\label{P} \psi= \frac{\sigma(\mu)}{\sqrt{2}} \Psi
\end{equation}
Then the image of $\mu$ is again a Gaussian measure, say
$\mu_{\rm{scal}},$ having the dispersion $\sigma^2(\mu_{\rm{scal}})=
2.$ Set $D=\rm{cov}\;\mu_{\rm{scal}}.$ In the two dimensional case $
D= \left(
\begin{array}{ll}
 1 & 0\\
0 & 1
\end{array}
 \right )$ and  $\rm{Tr}\; D=2.$ We now have:
\begin{equation}
\label{P1} <f>_\mu(\alpha) = \frac{\sigma^2(\mu)}{4}  \rm{Tr} \;D\;
f^{\prime \prime}(0) +O(\alpha^2).
\end{equation}
Thus
\begin{equation}
\label{P2} <f>_\mu(\alpha) = \frac{\sigma^2(\mu)}{2} \;j(D)\;
j(f^{\prime \prime}(0)) +O(\alpha^2).
\end{equation}
Finally, we note that in the two dimensional case: $j(D)=1.$ Thus we
obtain:
\begin{equation}
\label{P3} <f>_\mu(\alpha) = \frac{\sigma^2(\mu)}{2} \;j(f^{\prime
\prime}(0)) +O(\alpha^2),
\end{equation}
and hence (\ref{P0}).

\medskip

We recall that in the one dimensional quantum mechanics there is
just one ``density matrix'', namely, $\rho=1\in {\bf R}.$

It is convenient to consider the renormalization of averages by the
main term in the intensities of fluctuations of the coordinate and
momenta: $\langle f \rangle_\mu= \frac{<f>_\mu}{\alpha}.$ Then we
get:
\begin{equation}
\label{P0F} \langle f \rangle_\mu(\alpha) = j(f^{\prime\prime}(0)) +
O(\alpha).
\end{equation}

\section{Prequantum phase space -- multidimensional case}

States of systems are now represented by points $\psi=(q,p)\in
\Omega= Q\times P,$ where $Q=P={\bf R}^m.$ Here the
$q=(q_1,...,q_n)$ is the position and the $p=(p_1,...,p_n)$ is
momentum, so $\Omega$ denotes phase space. Let us consider the
canonical representation of the group $SO(2)$ in the phase space
$\Omega= Q\times P:$

\begin{equation}
\label{RER1} u=u_\theta= \left( \begin{array}{ll}
 \cos \theta \; I & - \sin \theta \; I\\
\sin \theta \; I & \cos \theta \; I
\end{array}
 \right ),
\end{equation}
where $I$ is the unit matrix from $M^{(r)}(m).$  The corresponding
group of ${\bf R}$-linear operators (or $2m\times 2m$ matrices) we
denote by the symbol $SO_m(2).$

 The classical model $N_{\rm{class}}=
(S_G^\alpha(\Omega \vert SO_m(2)), {\cal V}(\Omega \vert SO_m(2)))$
in defined in the same way as in the two dimensional case. A
Gaussian measure is invariant iff its covariance operator belongs to
the commutator $SO_m^\prime(2)=\{ A\in M^{(r)}(2m): [A, u]=0, u \in
SO_m(2)\}.$ If a smooth function $f$ is invariant then all its odd
derivatives equal to zero and the second derivative belong to the
$SO_m^\prime(2).$ A matrix $A\in SO_m^\prime(2)$ if it has the form
(\ref{U}), where $R, S\in M^{(r)}(m).$ In contrast to the two
dimensional case a symmetric matrix from $SO_m^\prime(2)$ can be
nondiagonal. It has the form (\ref{U}), where $R^*=R$ and $S^*=-S.$

There is  a natural map (generalizing the map $j: SO^\prime(2) \to
{\bf C})$ of the commutator $SO_m^\prime(2)$ onto the set of complex
matrices $M^{(c)}(m):$
\begin{equation} \label{ZX0G}
j: SO_m^\prime(2) \to M^{(c)}(m), z=j(A)= R+iS .
\end{equation}
This is the isomorphism of two rings. Symmetric matrices are mapped
onto symmetric matrices. Let us denote real and complex conjugations
by $*$ and $\star,$ respectively. We have $ (R+iS)^\star = R^* -i
S^*= R+iS. $
\medskip
We also remark that for a symmetric complex matrix:
\begin{equation} \label{ZXG}
\rm{Tr}\; j(A) = \rm{Tr}\;(R+iS)= \rm{Tr}\; R = \frac{1}{2} \;
\rm{Tr}\; A.
\end{equation}

\medskip

{\bf Lemma 8.1.} {\it Let $f \in {\cal V}(\Omega \vert SO_m(2))$ and
let $\mu \in S_G^\alpha(\Omega \vert SO_m(2)).$ Then
\begin{equation}
\label{P0Q} <f>_\mu(\alpha) = \alpha \; \rm{Tr} \rho \;
j(f^{\prime\prime}(0)) + O(\alpha^2),
\end{equation}
where $\rho \in {\cal D}^{(c)}(m).$}

{\bf Proof.} We make the scaling (\ref{P}) and get the
$\mu_{\rm{scal}}$ with $D=\rm{cov}\;\mu_{\rm{scal}},$ and $\rm{Tr}\;
D=2.$ We set $\rho= j(D),$  here $\rm{Tr}\; \rho= (\rm{Tr}\; D)/2=1$
and $\rho \in {\cal D}^{(c)}(m).$ Finally
\begin{equation}
\label{P2Q} <f>_\mu(\alpha) = \frac{\sigma^2(\mu)}{2} \;\rm{Tr}\;
j(D)\; j(f^{\prime \prime}(0)) +O(\alpha^2)
\end{equation}
implies (\ref{P0Q}).

\medskip

We now modify the classical$\to$ quantum projections, (\ref{KK1}),
(\ref{KK2}), to make them consistent with the complex structure:
\begin{equation}
\label{KK1B} T: S_{G}^\alpha(\Omega\vert SO_m(2)) \to {\cal
D}^{(c)}(m), \; \rho=T(\mu)= j(\rm{cov}\; \mu_{\rm{scal}});
\end{equation}
\begin{equation}
\label{KK2B} T: {\cal V}(\Omega \vert SO_m(2)) \to {\cal
L}_s^{(c)}(m),\; A=T(f)=j(f^{\prime\prime})(0)
\end{equation}

{\bf Theorem 8.1.} {\it The maps (\ref{KK1B}), (\ref{KK2B}) project
the classical statistical model $N_{\rm{class}}= (S_G^\alpha(\Omega
\vert SO_m(2)), {\cal V}(\Omega \vert SO_m(2)))$  onto the quantum
model $N_{\rm{quant}}^{(c)}= ({\cal D}^{(c)}(m), {\cal
L}_s^{(c)}(m))$  in such a way that classical and quantum averages
are coupled by the asymptotic equality:}
\begin{equation}
\label{KK3T} \langle f \rangle_\mu=  <T(f)>_{T(\mu)} + O(\alpha).
\end{equation}

\section{Prequantum phase space}

States of systems are now represented by points $\psi=(q,p)\in
\Omega= Q\times P,$ where $Q=P=H$ and $H$ is a real (separable
Hilbert space) with the scalar product $(\cdot, \cdot)$ and the
corresponding norm $\Vert \cdot \Vert.$ Here the $q\in H$ is the
position and the $p\in H$ is momentum, so $\Omega$ denotes phase
space. The real Hilbert space structure on $\Omega$ is given by the
scalar product
\begin{equation}
\label{SP} (\psi_1, \psi_2)= (q_1,q_2)+ (p_1,p_2).
\end{equation}
In physics $H=L_2({\bf R}^3)$ is the space of square integrable
functions. Thus both position and momentum are functions of $x\in
{\bf R}^3.$ These are simply classical fields. A point of such a
phase space is a classical vector field $\psi(x) =(q(x),p(x)).$

Let us consider the canonical representation of the group $SO(2)$ in
the phase space $\Omega= Q\times P,$ see (\ref{RER1}), where $I:H
\to H$ is the unit operator.  The corresponding group of continuous
${\bf R}$-linear operators we denote by the symbol $SO_H(2).$

The classical model
$$
N_{\rm{class}}= (S_G^\alpha(\Omega \vert SO_H(2)), {\cal V}(\Omega
\vert SO_H(2)))
$$
in defined in the same way as in the finite-dimensional case. We
just recall a few basic notions from theory of differentiable
functions and Gaussian measures on infinite-dimensional spaces.

Let $\mu$ be a $\sigma$-additive Gaussian measure on the
$\sigma$-field $F$ of Borel subsets of $\Omega,$ see [32]. This
measure is determined by its covariance operator $B: \Omega\to
\Omega$ and mean value $m\equiv m_\mu \in  \Omega.$ For example, $B$
and $m$ determines the Fourier transform of $\rho:$
$$
\tilde\rho (y)= \int_\Omega e^{i(y, \psi)} d\mu (\psi)=
e^{\frac{1}{2}(By, y) + i(m, y)}, \; y \in  \Omega.
$$
In what follows we restrict our considerations to Gaussian measures
with zero mean value $m=0,$ where $(m,y) = \int_\Omega (y, \psi)
d\mu(\psi)= 0$ for any $y \in \Omega.$ We recall that the covariance
operator $B\equiv \rm{cov} \; \mu$ is defined by
$$
(By_1, y_2)=\int_\Omega (y_1, \psi) (y_2, \psi) d\mu(\psi), y_1, y_2
\in \Omega ,
$$
and has the following properties: a). $B \geq 0,$ i.e., ($By, y)
\geq 0, y \in \Omega;$ b). $B$ is a bounded self-adjoint operator,
$B \in {\cal L}_{s}(\Omega );$ c). $B$ is a trace-class operator and
moreover
$${\rm Tr}\; B=\int_\Omega ||\psi||^2 d\mu(\psi).
$$
This is {\it dispersion} $\sigma^2(\mu)$ of the probability $\mu.$
Thus $\sigma^2(\mu)= {\rm Tr}\; B.$

We remark that the list of properties of the covariance operator of
a Gaussian measure differs from the list of properties of a von
Neumann density operator [4] only by one condition: $\rm{Tr} \; \rho
=1,$ for a density operator $\rho.$

We can easily find the Gaussian integral of a quadratic form (by
using expansion with respect to an orthonormal basis and using our
previous results on the finite-dimensional Gaussian integrals):
$\int_\Omega (A\psi, \psi) d\rho (\psi)={\rm Tr}\; BA,$ where $A\in
{\cal L}_s(\Omega).$

The differential calculus for maps $f: \Omega \to {\bf R}$ does not
differ so much from the differential calculus in the finite
dimensional case, $f: {\bf R}^n \to {\bf R}.$ Instead of the norm on
${\bf R}^n,$ one should use the norm on $\Omega.$ We consider so
called Frechet differentiability. Here a function $f$ is
differentiable if it can be represented as
$$
f(\psi_0 + \Delta \psi)= f(\psi_0) + f^\prime(\psi_0)(\Delta \psi) +
o(\Delta \psi),
$$
where $\lim_{\Vert \Delta \psi \Vert\to 0}
\frac{\Vert o(\Delta \psi)\Vert }{\Vert \Delta \psi \Vert} =0.$ Here
at each point $\psi$ the derivative $f^\prime(\psi)$ is a continuous
linear functional on $\Omega;$ so it can be identified with the
element $f^\prime(\psi)\in \Omega.$ Then we can define the second
derivative as the derivative of the map $\psi \to f^\prime(\psi)$
and so on. A map $f$ is differentiable $n$-times iff: $$ f(\psi_0 +
\Delta \psi)= f(\psi_0) + f^\prime(\psi_0)(\Delta \psi) +
\frac{1}{2}f^{\prime \prime}(\psi_0)(\Delta \psi, \Delta \psi) + ...
$$
$$
+\frac{1}{n!} f^{(n)}(\psi_0)(\Delta \psi, ..., \Delta \psi)+
o_n(\Delta \psi),$$ where $f^{(n)}(\psi_0)$ is a symmetric
continuous $n$-linear form on $\Omega$ and
$$
\lim_{\Vert \Delta \psi
\Vert\to 0} \frac{\Vert o_n(\Delta \psi)\Vert }{\Vert \Delta \psi
\Vert^n} =0.
$$
For us it is important that the second derivative
$f^{\prime\prime}(\psi_0)$ can be represented by a self-adjoint
operator $f^{\prime \prime}(\psi_0)(u,v)=(f^{\prime \prime}(\psi_0)
u, v), u, v \in \Omega.$ We remark that for $\psi_0=0$ we have:
$$
f(\psi)= f(0) + f^\prime(0)(\psi) + \frac{1}{2}f^{\prime
\prime}(0)(\psi, \psi) + ...+ \frac{1}{n!} f^{(n)}(0)(\psi,
...,\psi) + o_n(\psi).
$$
As in the finite-dimensional case the
reminder can be represented in the integral form.

A Gaussian measure is invariant iff its covariance operator belongs
to the commutator
$$
SO_H^\prime(2)=\{ A\in {\cal L}(\Omega): [A, u]=0, u \in SO_H(2)\}.
$$
If a smooth function $f: \Omega \to {\bf R}$ is
$SO_H^\prime(2)$-invariant then all its odd derivatives equal to
zero and the second derivative belong to the $SO_H^\prime(2).$ An
operator $A\in SO_H^\prime(2)$ if it has the form (\ref{U}), where
$R, S\in {\cal L}(H).$ And $A\in SO_H^\prime(2) \cap {\cal
L}_s(\Omega)$ if it has the form (\ref{U}), where $R^*=R$ and
$S^*=-S.$

Let us now consider the complexification of the Hilbert space $H:$
$H_c= H \oplus i H.$ We denote the algebra of bounded ${\bf
C}$-linear operators $A: H_c \to H_c$ by the symbol ${\cal L}(
H_c).$ The set of self-adjoint operators (with respect to the
complex scalar product) we denote by the symbol ${\cal L}_s(H_c).$

There is  a natural map (generalizing the map $j: SO_m^\prime(2) \to
{\bf C}^m)$ of the commutator $SO_H^\prime(2)$ onto the ${\cal L}(
H_c):$
\begin{equation} \label{ZX0GR}
j: SO_H^\prime(2) \to {\cal L}( H_c), z=j(A)= R+iS .
\end{equation}
This is the isomorphism of two rings. Self-adjoint operators (with
respect to the real scalar product) are mapped onto self-adjoint
operators (with respect to the complex scalar product). We also
remark that for a self-adjoint trace class operator $A\in
SO_H^\prime(2)$ the equality (\ref{ZXG}) coupling real and comlex
traces holds. In the same way as in the finite-dimensional case we
prove:

{\bf Lemma 9.1.} {\it Let $f \in {\cal V}(\Omega \vert SO_H(2))$ and
let $\mu \in S_G^\alpha(\Omega \vert SO_H(2)).$ Then the asymptotic
equality (\ref{P0Q}), where $\rho \in {\cal D}(H_c),$ holds.}

We now consider the infinite-dimensional generalization of  the
classical$\to$ quantum projections, (\ref{KK1B}), (\ref{KK2B})
\begin{equation}
\label{KK1D} T: S_{G}^\alpha(\Omega\vert SO_H(2)) \to {\cal D}(H_c),
\; \rho=T(\mu)= j(\rm{cov}\; \mu_{\rm{scal}});
\end{equation}
\begin{equation}
\label{KK2D} T: {\cal V}(\Omega \vert SO_H(2)) \to {\cal L}_s(H_c),
\; A=T(f)=j(f^{\prime\prime})(0)
\end{equation}
Lemma 9.1 implies:

{\bf Theorem 9.1.} {\it The maps (\ref{KK1D}), (\ref{KK2D}) project
the classical statistical model $N_{\rm{class}}= (S_G^\alpha(\Omega
\vert SO_H(2)), {\cal V}(\Omega \vert SO_H(2)))$  onto the quantum
model $N_{\rm{quant}}(H_c)= ({\cal D}(H_c), {\cal L}_s(H_c))$ in
such a way that classical and quantum averages are coupled by the
asymptotic equality (\ref{KK3T}).}

We remark that the idea that the quantum averages can be coupled to
integration with respect to the $\psi$-function was discussed in a
number of papers, see, e.g., \cite{Bach1}-- \cite{Bach3} and \cite{SG} (so called GAP-measures)
as well as extended literature in the last paper. The main distinguishing feature of our
approach is elaboration of technique of asymptotic expansion with
respect to a small parameter, namely the dispersion of prequantum
fluctuations. Comparing with \cite{SG} we also mention that we
(as well as Bach \cite{Bach1}-- \cite{Bach3}) consider the linear space integration and not integration 
over the unit sphere. This is not simply a technical
deviation, but it implies a totally new viewpoint to quantum pure
states, see the next section.

\section{Gaussian measures corresponding to pure quantum states}

We now generalize considerations of section 6 to the complex
infinite-dimensional case. Let $\Psi$ be a pure quantum state: $\Psi
\in H_c, \Vert \Psi \Vert =1.$ We define a Gaussian measure
$\mu_\Psi$ which is concentrated on the one dimensional complex
space (so the real plane) $\Pi_\Psi=\{\psi \in H_C: \psi =c \Psi, c
\in{\bf C}\}:$ the average of the $\mu_{\Psi}$ is equal to zero and the complexification
of its covariance operator:
$$
j(\rho_{\mu_{\Psi}})= \alpha \Psi \otimes \Psi.
$$
The following facts about $\mu_{\Psi}$ can be obtained through
direct computations and Theorem 9.1:

\medskip

{\bf Proposition 10.1.} {\it For any pure quantum state $\Psi,$ we
have:

a). $\mu_{\Psi} \in S_G^\alpha(\Omega \vert SO_H(2));$

b). $T(\mu_{\Psi})= \Psi \otimes \Psi;$

c). $\langle f \rangle_\mu=  <j(f^{\prime \prime}(0)) \Psi,\Psi>  +
O(\alpha), \; f \in {\cal V}(\Omega \vert SO_H(2)).$}

\section{Hamilton-Schr\"odinger dynamics}

States of systems with the infinite number of degrees of freedom -
classical fields -- are represented by points $\psi=(q, p) \in
\Omega;$ evolution of a state is described by the Hamiltonian
equations. We consider a quadratic Hamilton function: ${\cal H}(q,
p)=\frac{1}{2} ({\bf H} \psi,\psi),$ where ${\bf H}: \Omega \to
\Omega$ is an arbitrary symmetric (bounded) operator; the
Hamiltonian equations have the form:
$$
\dot q= {\bf H}_{21}q + {\bf
H}_{22} p, \; \; \dot p=-( {\bf H}_{11}q +{\bf H}_{12}p),
$$
or
\begin{equation} \label{YC} \dot \psi= \left(
\begin{array}{ll}
\dot q\\
\dot p
\end{array}
\right )=J{\bf H} \psi
\end{equation}
Thus quadratic Hamilton functions induce linear Hamilton equations.
From (\ref{YC}) we get
$$
\psi(t)= U_t \psi, \; \; \mbox{where} \; U_t=e^{J {\bf H} t}.
$$
The map $U_t\psi$ is a linear Hamiltonian flow on the phase space
$\Omega.$ Let us consider a self-adjoint operator ${\bf H} \in
SO_H^\prime(2)$: ${\bf H}= \left(
\begin{array}{ll}
R&T\\
-T&R
\end{array}
\right).$ This operator defines the quadratic  Hamilton function
$$
{\cal H}(q, p)=\frac{1}{2}[(R p, p) + 2 (Tp, q) + (Rq, q)],
$$
where the operator $R$ is symmetric and the operator $T$ is skew
symmetric. Corresponding Hamiltonian equations have the form
\begin{equation}
\label{CZ1}\dot q=Rp-Tq, \;  \dot p=-(Rq + Tp).\end{equation} We
point out that for a $SO_H(2)$-invariant Hamilton function, the
Hamiltonian flow $U_t \in SO_H^\prime(2).$ By considering the
complex structure on the infinite-dimensional phase space $\Omega$
we write the Hamiltonian equations (\ref{YC}) in the form of the
Sch\"odinger equation on $H_c:$
\begin{equation}
\label{YV} i  \frac{d \psi}{d t} = {\bf H} \psi .
\end{equation} Its solution has
the following complex representation: $\psi(t)=U_t \psi, \; \;
U_t=e^{-i{\bf H} t}.$ We consider the Planck system of units in that
$h=1.$ This is {\it the complex representation of flows
corresponding to quadratic $SO_H(2)$-invariant Hamilton functions.}

By choosing $H=L_2( {\bf R}^n)$ we see that the interpretation of
the solution of this equation coincides with the original
interpretation of Schr\"odinger -- this is a classical field
$$\psi(t,x)=(q(t,x), p(t,x).$$

\medskip

{\bf Example 11.1.} Let us consider an important class of Hamilton
functions
\begin{equation}
\label{HF} {\cal H} (q, p)=\frac{1}{2}[(Rp, p)+(Rq, q)],
\end{equation}
where $R$ is a symmetric operator. The corresponding Hamiltonian
equations have the form:
\begin{equation}
\label{HF1} \dot q=Rp, \; \dot p=-Rq.
\end{equation}
We now choose $H=L_2({\bf R}^3),$ so $q(x)$ and $p(x)$ are
components of the vector-field $\psi(x)=(q(x), p(x)).$ We can call
fields $q(x)$ and $p(x)$ {\it mutually inducing.} The field $p(x)$
induces dynamics of the field $q(x)$ and vice versa, cf. with
electric and magnetic components, $q(x)=E(x)$ and $p(x)=B(x),$ of
the electromagnetic field, cf. Einstein and Infeld \cite{EI}, p. 148:
{\small ``Every change of an electric field produces a magnetic
field; every change of this magnetic field produces an electric
field;  every change of ..., and so on.''} We can write the form
(\ref{HF}) as
\begin{equation}
\label{HF2} {\cal H} (q, p)=\frac{1}{2} \int_{{\bf R}^6} R(x, y) [q
(x) q(y) + p(x) p(y)] dx dy
\end{equation}
or
\begin{equation}
\label{HF3} {\cal H}(\psi)=\frac{1}{2} \int_{{\bf R}^6} R(x, y) \psi
(x) \bar{\psi} (y) dx dy ,
\end{equation}
where $R (x, y)=R(y, x)$ is in general a distribution on ${\bf
R}^6.$ We call such a  kernel $R(x, y)$  a {\it self-interaction
potential} for the background field $\psi(x)=(q(x), p(x)).$We point
out that $R(x, y)$ induces a self-interaction of each component of
the $\psi(x),$ but there is no cross-interaction between components
$q(x)$ and $p(x)$ of the vector-field $\psi(x).$

\section{Invariant Gaussian measures of the Hamilton-Schr\"odinger dynamics and stationary pure states}

All Gaussian measures considered in this section are supposed to be
$SO_H(2)$-invariant. As we have seen, In our approach so called pure
states $\Psi, ||\Psi||=1,$ are labels for Gaussian measures
concentrated on one dimensional (complex) subspaces $\Omega_\Psi$ of
the infinite-dimensional phase-space $\Omega.$ In this section we
study the case of so called {\it stationary (pure) states} in more
detail. The $\alpha$-scaling does not play any role in present
considerations. Therefore we shall not take it into account. We
consider a pure state $\Psi, ||\Psi||=1,$ as the label for the
Gaussian measure  $\nu_\Psi$ having the zero mean value and the
complexification of the covariance operator
$$
j(\rho_{\nu_\Psi})=\Psi \otimes \Psi.
$$

\medskip

{\bf Theorem 12.1.} {\it Let $\nu$ be a Gaussian measure (with zero
mean value) concentrated on the one-dimensional (complex) subspace
corresponding to a normalized vector $\Psi$. Then $\nu$ is invariant
with respect to the unitary dynamics $U_t=e^{-it {\bf H}},$ where
${\bf H}: \Omega \to \Omega$ is a bounded self-adjoint operator, iff
$\Psi$ is an eigenvector of $\bf H.$}

{\bf Proof.} A). Let ${\bf H} \Psi= \lambda \Psi.$ The Gaussian
measure  $U_t^* \nu$ has the complexification of the covariance
operator
$$
j(\rho_t)= U_t (\Psi \otimes \Psi) U_t^*= U_t \Psi
\otimes U_t \Psi= e^{-it\lambda} \Psi \otimes e^{-it \lambda} \Psi=
\Psi \otimes \Psi.
$$
Since all measures under consideration are
Gaussian, this implies that  $U_t^* \nu= \nu.$ Thus $\nu$ is an
invariant measure.

B). Let $U_t^* \nu=\nu$ and $\nu=\nu_\Psi$ for some $\Psi,
||\Psi||=1.$ We have that $U_t \Psi \otimes U_t\Psi=\Psi \otimes
\Psi.$ Thus, for any $\psi_1, \psi_2 \in \Omega,$ we have
\begin{equation}
\label{COVZ}
  \langle \psi_1,U_t \Psi \rangle  \langle U_t \Psi, \psi_2 \rangle=  \langle \psi_1,
\Psi \rangle  \langle \Psi, \psi_2 \rangle.
\end{equation}
Let us set $\psi_2=\Psi.$ We obtain: $  \langle \psi_1,
\overline{c(t)} U_t \Psi \rangle=  \langle \psi_1, \Psi \rangle,$
where $c(t)=  \langle U_t \Psi, \Psi \rangle. $ Thus
$\overline{c(t)} U_t \Psi=\Psi.$We point out that
$c(0)=||\Psi||^2=1.$ Thus $ \overline{c^\prime (0)} \Psi -i {\bf H}
\Psi=0,$ or ${\bf H} \Psi=-i \overline{c^\prime (0)} \Psi.$ Thus
$\Psi$ is an eigenvector of $\bf H$ with the eigenvalue $-i
\overline{c^\prime (0)}.$ We remark that $c^\prime (0)=-i
 \langle {\bf H} \Psi, \Psi \rangle; $ so $ \overline{c^\prime (0)}=i
  \langle {\bf H} \Psi, \Psi \rangle.$ Hence, $\lambda=-i  \overline{c^\prime
(0)}=  \langle {\bf H}, \Psi, \Psi \rangle.$

\medskip

{\bf Conclusion.} {\it {In PCSFT stationary states of the quantum
Hamiltonian (represented by a bounded self-adjoint operator $\bf H)$
are labels for Gaussian one-dimensional measures (with the zero mean
value) that are invariant with respect to the Schr\"odinger dynamics
$U_t=e^{-it {\bf H}}$.}}

\medskip

We now describe all possible Gaussian measures which are
$U_t$-invariant.

{\bf Theorem 12.2.} {\it {Let $\bf H$ be a bounded self-adjoint
operator with purely discrete nondegenerate spectrum: ${\bf H}
\Psi_k=\lambda_k \Psi_k,$ so $\{\Psi_k\}$ is an orthonormal basis
consisting of eigenvectors of $\bf H.$ Then any $U_t$-invariant
Gaussian measure $\nu$ (with the zero mean value) has the
complexification of the covariance operator:

\begin{equation}
\label{COV} j(\rho)=\sum_{k=1}^\infty c_k \Psi_k \otimes \Psi_k,
c_k\geq 0,
\end{equation}

and vice versa.}}

{\bf Proof.} A). Let $j(\rho)$ has the form (\ref{COV}). Then
\begin{equation}
\label{ZEEE} j(\rho_{U_t^* \nu})= U_t j(\rho) U_t^*=
\sum_{k=1}^\infty c_k e^{-i \lambda_k t} \Psi_k \otimes
e^{-i\lambda_k t} \Psi_k = j(\rho).
\end{equation}
Since measures are Gaussian, this implies that $U_t^* \nu=\nu$ for
any $t.$

B). Let $U_t^* \nu=\nu$ for any $t.$ We remark that the
complexification of any covariance operator $\rho$ can be
represented in the form:
\begin{equation}
\label{ZEEE1} j(\rho) =\sum_{k=1}^\infty   \langle j(\rho) \Psi_k,
\Psi_k \rangle \Psi_k \otimes \Psi_k + \sum_{k \ne l} \langle
j(\rho) \Psi_k, \Psi_l \rangle \Psi_k \otimes \Psi_l.
\end{equation}
We shall show that $\langle j(\rho) \Psi_k, \Psi_j \rangle=0$ for $k
\ne j.$ Denote the operator corresponding to $\sum_{k \ne j}$ by Z.
We have
\begin{equation}
\label{ZEEE2} \langle U_t Z U_t \psi_1, \psi_2 \rangle= \sum_{k \ne
m} \langle j(\rho) \Psi_k, \Psi_m \rangle e^{it(\lambda_m -
\lambda_k)} \langle \Psi_k, \psi_2 \rangle   \langle \psi_1, \Psi_m
\rangle=
 \langle Z \psi_1, \psi_2 \rangle.
\end{equation}
Set $\psi_1=\Psi_j, \psi_2= \Psi_k.$ Then
\begin{equation}
\label{ZEEE3} \langle U_t Z U_t^* \Psi_m, \Psi_k \rangle= \langle
j(\rho) \Psi_k, \Psi_m \rangle e^{it(\lambda_m - \lambda_k)}=
\langle j(\rho) \Psi_k, \Psi_m \rangle.
\end{equation}
Thus $\langle j(\rho) \Psi_k, \Psi_m \rangle=0, k \ne m.$

\section{Stability of hydrogen atom}

As we have seen, in PCSFT so called stationary (pure) states of
quantum mechanics can play the role of labels for Gaussian measures (which are
$SO_H(2)$-invariant and have zero mean value) that are
$U_t$-invariant. We now apply our standard $\alpha$-scaling argument
and we see that a stationary state $\Psi$ is a label for the
Gaussian measure $\mu_\Psi$ with $j(\rho_{\mu_\Psi})=\alpha \Psi
\otimes \Psi.$ This measure is concentrated on one-dimensional
(complex) subspace $\Pi_\Psi$ of phase space $\Omega.$ Therefore
{\it each realization of an element of the Gaussian ensemble of
classical fields corresponding to the statistical state $\mu_\Psi$
gives us the field of the shape $\Psi(x),$ but magnitudes of these
fields vary from one realization to another.} But by the well known
Chebyshov inequality probability that ${\cal E}(\Psi)=\int_{{\bf
R}^3}|\Psi(x)|^2 dx$ is large is negligibly small.

Thus in the stationary state we have {\it Gaussian fluctuations of
very small magnitudes of the same shape} $\Psi(x).$ In PCSFT a
stationary quantum state can not be identified with a stationary
classical field, but only with an ensemble of fields having the same
shape $\Psi(x).$ Let us now compare descriptions of dynamics of
electron in hydrogen atom given by quantum mechanics and our
prequantum field theory.

In quantum mechanics stationary bound states of hydrogen atom are of
the form:  \begin{equation} \label{UI} \Psi_{nlm}(r, \theta,
\phi)=c_{n,l} R^l L_{n + l}^{2l + 1} (R) e^{-R/2} Y_l^m (\theta,
\phi),\end{equation} where $R=\frac{2r}{n a_0},$ and
$a_0=\frac{h^2}{\mu e^2}$ is a characteristic length for the atom
(Bohr radius). We are mainly interested in the presence of the
component $e^{-R/2}.$

In PCSFT this stationary bound state is nothing else, but the label
for the Gaussian measure $\rho_{\Psi_{nlm}}$ which is concentrated
on the subspace $\Omega_{\Psi_{nlm}}.$ Thus PCSFT says that
``electron in atom'' is nothing else than Gaussian fluctuations of a
certain classical field, namely the field  $\Psi_{nlm} (r, \theta,
\phi):$
\begin{equation}
\label{FL} \psi_{nlm} (r, \theta, \phi; \psi)=\gamma (\psi)
\Psi_{nlm} (r, \theta, \phi),
\end{equation}
where $\gamma(\psi)$ is the C-valued Gaussian random variable:
$E\gamma=0, E|\gamma|^2=\alpha.$

The intensity of the field $\Psi_{nlm} (r, \theta, \phi, \psi)$
varies, but the shape is the same. Therefore this random field does
not produce any significant effect for large $R$ (since $e^{-R/2}$
eliminates such effects).

Thus in PCSFT the hydrogen atom stable, since the prequantum random
fields $\psi_{nlm} (r, \theta, \phi; \psi)$ have a special shape
(decreasing exponentially $R \to \infty).$

\section{Appendixes}

\subsection{Classical representation for spin operators}

The Pauli matrices are a set of $2 \times 2$ complex Hermitian and
unitary matrices.  They are: $\sigma_1 = \begin{pmatrix} 0&1\\
1&0 \end{pmatrix},$ $ \sigma_2 =\begin{pmatrix} 0&-i\\ i&0
\end{pmatrix},$ $ \sigma_3 =
\begin{pmatrix} 1&0\\ 0&-1 \end{pmatrix}.$ Let $H_c= C^2$  with the
complex coordinates $z=(z_1,z_2),\; z_j=q_j+ip_j, j=1,2,$ and
$\Omega= {\bf R}^2 \times {\bf R}^2$ with the real coordinates
$\omega=(q_1,q_2, p_1,p_2).$ We consider spin operators: $
\sigma(a)= \sum_{j}^3 a_j \sigma_j: {\bf C}^2 \to {\bf C}^2, \;
a=(a_1,a_2, a_3).$ Let us consider real matrices
$\sigma_j^{(r)}=j^{-1}(\sigma_j):$
$$
\sigma_1^{(r)} = \begin{pmatrix} \sigma_1 & 0\\
0 & \sigma_1 \end{pmatrix} =  \begin{pmatrix} 0 & 1 & 0 & 0\\
1 & 0 & 0 & 0\\
0 & 0 & 0 & 1\\
0 & 0 & 1 & 0
\end{pmatrix}, \; \sigma_2^{(r)} = \begin{pmatrix} 0 & -i \sigma_2\\
i \sigma_2 & 0 \end{pmatrix} =  \begin{pmatrix} 0 & 0 & 0 & -1\\
0 & 0 & 1 & 0\\
0 & 1 & 0 & 0\\
-1 & 0 & 0 & 0
\end{pmatrix},
$$
$$
\sigma_3^{(r)} = \begin{pmatrix} \sigma_3 & 0\\
0 & \sigma_3 \end{pmatrix} =  \begin{pmatrix} 1 & 0 & 0 & 0\\
0 & -1 & 0 & 0\\
0 & 0 & 1 & 0\\
0 & 0 & 0 & -1
\end{pmatrix}.
$$
We remark that these are not Dirac matrices. We set
$$
\sigma^{(r)}(a)\equiv j^{-1}(\sigma(a))= \sum_{j}^3 a_j
\sigma_j^{(r)}: {\bf R}^4 \to {\bf R}^4, \; a=(a_1,a_2, a_3).
$$
and consider {\it classical random spin-variables}: $ f_a(\omega)=
\frac{1}{2} (\sigma^{(r)}(a) \omega, \omega). $ Then $T(f_a)=
\sigma(a)$ and for any $SO_2(2)$-invariant Gaussian measure $\mu$ on
$\Omega= {\bf R}^2 \times {\bf R}^2$ with dispersion $\alpha+
O(\alpha)$ we have: $ \frac{1}{\alpha} \int_{{\bf R}^4} f_a(\omega)
d\mu(\omega)= \rm{Tr} \; j(\rho) \sigma(a) + O(\alpha), $ where
$\rho$ is the covariance operator of $\sqrt{\alpha}$-scaling of the
Gaussian measure $\mu.$ For example, $$ \frac{1}{\alpha} \int_{{\bf
R}^4} (q_1 q_2 + p_1 p_2) d\mu(q_1, q_2, p_1, p_2)= \rm{Tr}\;
j(\rho) \sigma_1 + O(\alpha),
$$
$$
\frac{1}{\alpha}  \int_{{\bf R}^4} (p_1 q_2 - p_2 q_1) d\mu(q_1,
q_2, p_1, p_2)= \rm{Tr} \; j(\rho) \sigma_2 + O(\alpha),
$$
$$
\frac{1}{\alpha}  \int_{{\bf R}^4} (q_1^2 - q_2^2 + p_1^2 - p_2^2)
d\mu(q_1, q_2, p_1, p_2)= \rm{Tr} \; j(\rho) \sigma_3 + O(\alpha).
$$
We also have: $ \frac{1}{\alpha}  \int_{{\bf R}^4} (q_1^2 + q_2^2 +
p_1^2 + p_2^2) d\mu(q_1, q_2, p_1, p_2)= \rm{Tr} \; j(\rho) I +
O(\alpha)= \rm{Tr} \; j(\rho) + O(\alpha).
$
By introducing vectors $\omega_1= (q_1, p_1)$ and $\omega_2= (q_2,
p_2)$ we rewrite these asymptotic equalities in shorter way: $$
\frac{1}{\alpha}  \int_{{\bf R}^4} (\omega_1, \omega_2)
d\mu(\omega_1, \omega_2)= \rm{Tr}\;  j(\rho) \sigma_1 + O(\alpha),$$
$$\frac{1}{\alpha} \int_{{\bf R}^4} (J \omega_1, \omega_2)
d\mu(\omega_1, \omega_2)= \rm{Tr} \; j(\rho) \sigma_2 + O(\alpha),$$
where $J$ is the symplectic operator and, finally, $$
\frac{1}{\alpha}  \int_{{\bf R}^4} (\Vert \omega_1 \Vert^2 - \Vert
\omega_2 \Vert^2) d\mu(\omega_1, \omega_2)= \rm{Tr} \; j(\rho)
\sigma_3 + O(\alpha).$$

Let us now consider Gaussian measures on $\Omega= {\bf R}^4$
corresponding to pure quantum states. These are singular Gaussian
measures which are concentrated on $SO_2(2)$-invariant planes in
${\bf R}^4.$ To determine such a measure, we should find its
covariation operator.

\medskip

{\bf Proposition 14.1.} {\it Let $\Psi= u+ iv, u=(u_1, u_2)\in {\bf R}^2
, v=(v_1,v_2) \in {\bf R}^2$ be a pure quantum state and let
$\rho_\Psi=\Psi\times \Psi.$ Then $T^{-1}(\rho_\Psi) = \mu_\Psi,$
where the Gaussian measure $\mu_\Psi$ has the covariation operator
$B_\Psi=  \alpha D_\Psi$ for
$$
D_\Psi= \begin{pmatrix} \Vert g_1 \Vert^2 & (g_1,g_2) & 0 &
(Jg_1,g_2)\\  \\
(g_1,g_2) & \Vert g_2 \Vert^2 & (g_1,Jg_2) & 0\\
 \\
0 & (g_1, Jg_2) & \Vert g_1 \Vert^2 & (g_1,g_2)\\
 \\
(Jg_1,g_2) & 0 & (g_1,g_2) & \Vert g_2 \Vert^2
\end{pmatrix}.
$$
Here $g_1=(u_1,v_1)$ and $g_2=(u_2,v_2)$ are variables which
are conjugate to $\omega_1=(q_1,p_1)$ and $\omega_2=(q_2,p_2).$}

{\bf Proof.} The real space realization of $\rho_\Psi$ is given by
the operator:
\[j^{-1}(\rho_\Psi) =  \left( \begin{array}{ll}
 u \otimes u + v \otimes v & v \otimes u-u \otimes v\\
 u \otimes v-v \otimes u & u \otimes u + v \otimes v
 \end{array}
 \right ).
 \]
We have in the chosen system of coordinates on the phase space:
$$u \otimes u = \begin{pmatrix} u_1^2 & u_1u_2\\
u_1u_2 & u_2^2 \end{pmatrix}, \; v \otimes v = \begin{pmatrix} \\
v_1^2 & v_1v_2\\ v_1v_2 & v_2^2 \end{pmatrix}, $$ $$ v \otimes u= \begin{pmatrix} u_1 v_1 & u_2 v_1\\
u_1 v_2 & u_2 v_2\end{pmatrix},u \otimes v= \begin{pmatrix} v_1 u_1 & v_2 u_1\\
v_1 u_2 & v_2 u_2\end{pmatrix}.$$ Hence:
$$u \otimes u + v \otimes v = \begin{pmatrix} u_1^2 +v_1^2 & u_1u_2 + v_1v_2\\
u_1u_2+ v_1v_2 & u_2^2  +v_2^2 \end{pmatrix},$$ $$ v \otimes u-u
\otimes v= \begin{pmatrix} 0 & u_2 v_1- v_2 u_1\\
u_1 v_2 - v_1 u_2 & 0\end{pmatrix}.$$

\medskip

To illustrate better correspondence between real and complex state
spaces, we now show directly that $j(D_\Psi)= \Psi \otimes \Psi$ for
$D_\Psi$ given by this Proposition. We have
$$
j(D_\Psi)=\begin{pmatrix} \Vert g_1 \Vert^2 & (g_1,g_2) \\
(g_1,g_2) & \Vert g_2 \Vert^2 &
\end{pmatrix} + i \begin{pmatrix}   0 &
(Jg_1,g_2)\\
 (g_1,Jg_2) & 0
\end{pmatrix}.
$$
This operator acts to a complex vector $z=(z_1, z_2)$ in the
following way:
$$
z_1^\prime\equiv (j(D_\Psi)z)_1= \Vert g_1 \Vert^2 z_1+ [(g_1,g_2)+
i (Jg_1,g_2)]z_2,
$$
$$
z_2^\prime\equiv (j(D_\Psi)z)_2=  [(g_1,g_2)+ i(g_1,Jg_2)]z_1 +
\Vert g_2 \Vert^2 z_2.
$$
On the other hand, $\Psi \otimes \Psi(z)= <z, \Psi>\Psi=
(z_1\overline{\Psi}_1+ z_2\overline{\Psi}_2) \Psi.$ Here
$$z_1^\prime= (u_1 -i v_1)(u_1 + i v_1)z_1 + (u_2 - i v_2)(u_1 + i
v_1)z_2,$$
$$z_2^\prime= (u_1 -i v_1)(u_2 + i v_2)z_1 + (u_2 - i v_2)(u_2 + i
v_2)z_2.$$ Thus
$$
z_1^\prime=(u_1^2+v_1^2) z_1 +[(u_1 u_2 + v_1 v_2) +i (u_2 v_1 - u_1
v_2)] z_2,
$$
$$
z_2^\prime=(u_2^2+v_2^2) z_2 +[(u_1 u_2 + v_1 v_2) +i (u_1 v_2 -u_2
v_1)] z_2.
$$

Let us consider Gaussian measures corresponding to pure states for
spin up and spin down, $\vert 1>=\begin{pmatrix} 1\\
 0\end{pmatrix}$ and $\vert 0>= \begin{pmatrix} 0\\
 1\end{pmatrix}.$ For vector $\vert 1>,$ we have: $u_1=1, u_2=v_1=v_2=0.$
Thus $D_{\vert 1>}= \begin{pmatrix} 1 & 0 & 0 & 0\\
0 & 0 & 0 & 0\\
0 & 0 & 1 & 0\\
0 & 0 & 0 & 0
\end{pmatrix}$  and the Fourier transform of the measure $\mu_{\vert 1>}$ is
given by:
$$
\tilde{\mu}_{\vert 1>}(\xi_1, \xi_2, \eta_1,  \eta_2)=
e^{-\frac{\alpha}{2}(\xi_1^2 + \eta_1^2)}.
$$
This is the standard Gaussian measure on the plane $q_2=0, p_2=0$
having the density: $ d \mu_{\vert 1>}(q_1,p_1)= \frac{1}{2 \pi
\alpha} e^{-\frac{1}{2 \alpha}(q_1^2+p_1^2)}.$
 In the same way $\mu_{\vert 0>}$ is the standard Gaussian measure on the plane $q_1=0, p_1=0$
having the density: $ d \mu_{\vert 0>}(q_2,p_2)= \frac{1}{2 \pi
\alpha} e^{-\frac{1}{2 \alpha}(q_2^2+p_2^2)}.$ Let us now consider
the Gaussian measure corresponding to superposition of spin up and
spin down states:
$$\Psi_\theta=\frac{1}{\sqrt{2}} (\vert 0> + e^{i \theta}
\vert 1>).$$ Here $u_1= \cos \theta, v_1=  \sin \theta.$ Hence
$$
D_{\Psi_\theta}= \begin{pmatrix} \cos^2 \theta & \cos \theta \sin \theta & 0 & 0\\
\cos \theta \sin \theta & \sin^2 \theta & 0 & 0\\
0 & 0 & \cos^2 \theta & \cos \theta \sin \theta\\
0 & 0 & \cos \theta \sin \theta & \sin^2 \theta
\end{pmatrix}
$$
and the Fourier transform of $\mu_{\Psi_\theta}$ is given by
$$
\tilde{\mu}_{\Psi_\theta}(\xi_1, \xi_2, \eta_1,  \eta_2)=
e^{-\frac{\alpha}{2}[(\cos\theta\xi_1 + \sin\theta \xi_2)^2 + (\cos
\theta \eta_1 + \sin \theta \eta_2)^2]}.
$$
Thus pure states $\Psi_\theta$ correspond to the standard Gaussian
measures concentrated on planes obtained by rotations.

\subsection{Comparing with no-go theorems of von Neumann,
Cohen-Specker and Bell}

There are no-go theorems for mathematical attempts to have a map
from classical variables to quantum operators which preserves
statistics, e.g., theorems of von Neumann, Cohen-Specker and Bell,
see \cite{VN}, \cite{Bell}--\cite{Sh}. The no-go theorems say: No such map exists. In this paper
we constructed such a map. What goes?

Our construction does not contradict to known no-go theorems, since
our map $T$ does not satisfy some conditions of those theorems. An
important condition in all such theorems is that the {\it range of
values} of a classical variable $f$ should coincide with the
spectrum of the corresponding quantum operator $T(f)$ -- ``the range
of values postulate.'' This postulate is violated in our framework.
As we have seen, the classical spin variables are continuous and the
quantum spin operators have discrete spectrum. Nevertheless,
classical averages can be approximated by quantum. Our prequantum
classical statistical model is not about observations, but about
ontic reality (reality as it is when nobody looks at it).

Henry Stapp pointed out \cite{HS}: ``The problem, basically, is that to apply
quantum theory, one must divide the fundamentally undefined physical
world into two idealized parts, the observed and observing system,
but {\it the theory gives no adequate description of connection
between these two parts.} The probability function is a function of
degrees of freedom of the microscopic observed system, whereas the
probabilities it defines are probabilities of responses of
macroscopic measuring devices, and these responses are described in
terms of quite different degrees of freedom.'' Since we do know yet from physics  so
much about features of classical $\to$ quantum correspondence map $T,$
we have the freedom to change some conditions which were postulated
in the known no-go theorems -- for example, the range of values
condition. Rejection of this assumption is quite natural, since, as
was pointed by Stapp, a classical variable $f$ and its quantum
counterpart $T(f)$ depend on completely different degrees of
freedom.

\subsection{Is prequantum classical statistical field theory nonlocal?}

As we have seen, PCSFT does not contradict to the known no-go theorems,
in particular, to Bell's theorem. Therefore this theory might be  local. 
However, it is not easy to formulate the problem of
locality/nonlocality in the PCSTF-framework. It is not about
observations. Thus we could not apply Bell's approach \cite{Bell}--\cite{Sh} to locality as
locality of observations. On the other hand, on the ontic level
PCSTF operates not with particles, but with fields. At the first
sight, such a theory is nonlocal by its definition, since fields are
not localized. But in field theory there was established a different
viewpoint to locality and we know that both classical and quantum
field theories are local. To formulate the problem of locality for
PCSTF, we should proceed in the same way. Therefore we should
develop a relativistic version of PCSTF. There are some technical
and even ideological problems. As we know, relativistic quantum
mechanics is not a well established theory (at least this is a
rather common opinion). Thus it is meaningless  to develop a
relativistic variant of PCSTF  which would reproduce relativistic
quantum mechanics. The most natural way of development is  to
construct a kind of PCSTF  not for quantum mechanics, but for
quantum field theory and study the problem of locality in such a
framework. It is an interesting and complicated problem which will
be studied in coming papers of the author.


\begin{thebibliography}{99}


\bibitem{ES} E. Schr\"odinger,  {\it Philosophy and the Birth of Quantum
Mechanics.} Edited by M. Bitbol, O. Darrigol, Editions Frontieres,
Gif-sur-Yvette (1992); especially the paper of S. D'Agostino,
Continuity and completeness in physical theory: Schr\"odinger's
return to the wave interpretation of quantum mechanics in the
1950's, pp. 339-360.

\bibitem{PAM} P. A. M.  Dirac, The Principles of Quantum Mechanics, Oxford
Univ. Press (1930).

\bibitem{H}  W. Heisenberg, Physical principles of quantum theory, Chicago
Univ. Press (1930).

\bibitem{VN} J. von Neumann, {\it Mathematical foundations of quantum mechanics,}
Princeton Univ. Press, Princeton, N.J. (1955).

\bibitem{KH} A. Yu. Khrennikov, editor, Proc. Conf. {\it Foundations of
Probability and Physics,} Ser.  Quantum Probability and White Noise
Analysis, {\bf 13}, WSP, Singapore, 2001; Proc. Conf. {\it Quantum
Theory: Reconsideration of Foundations.} Ser. Math. Modeling , {\bf
2}, V\"axj\"o Univ. Press, V\"axj\"o, 2002; Proc.Conf. {\it
Foundations of Probability and Physics-2,} Ser. Math. Modeling, {\bf
5}, V\"axj\"o Univ. Press,  V\"axj\"o, 2003; Proc. Conf. {\it
Foundations of Probability and Physics-3,} American Institute of
Physics, Ser. Conf. Proc., {\bf 750}, Melville, NY, 2005.

\bibitem{11}   D. Bohm,  {\it Quantum theory,} Englewood Cliffs, New-Jersey,
Prentice-Hall (1951).

\bibitem{12}   P. Holland, {\it The quantum theory of motion,} Cambridge,
Cambridge Univ. Press  (1993).


\bibitem{4}   E. Nelson, {\it Quantum fluctuation,} Princeton: Princeton
Univ. Press (1985).


\bibitem{2}    L. De la Pena and   A. M. Cetto, {\it The Quantum Dice: An
Introduction to Stochastic Electrodynamics,} Dordrecht, Kluwer
(1996)


\bibitem{6}    G. `t Hooft,  Quantum Mechanics and Determinism, {\it Preprint}
hep-th/0105105.

\bibitem{KH1} A. Yu. Khrennikov, A pre-quantum classical statistical model with
infinite-dimensional phase space. {\it J. Phys. A: Math. Gen.}, {\bf
38}, 9051-9073 (2005).

\bibitem{KH2} A. Yu. Khrennikov, Generalizations of quantum mechanics induced
by Classical Statistical Field Theory. {\it Found. Phys. Letters},
{\bf 18}, 637-650 (2006).

\bibitem{EV} E. Ventzel, {\it Theory of probability,} Fizmatlit, Moscow,
1958.

\bibitem{BL} L. E. Ballentine, {\it Quantum mechanics,} Englewood Cliffs, New
Jersey, 1989.

\bibitem{KH3} A. Yu.  Khrennikov, Symplectic geometry on the Hilbert phase
space and foundations of quantum mechanics. Proc. of Conference
Modeling of Wave Phenomena, eds. B. Nilsson and L. Fishman, Vaxjo,
Sweden. AIP Conf. Proc., 634, Amer. Inst. Phys., Melville, NY, pp.
324-343(2006).


\bibitem{Bach1}  A.  Bach,   {\it J. Math. Phys.} {\bf 14} 125(1981).

\bibitem{Bach2}   A. Bach,   {\it Phys. Lett. A} {\bf 73} 287
(1979).

\bibitem{Bach3}   A. Bach, {\it J. Math. Phys.} {\bf 21} 789 (1980).

\bibitem{SG} S. Goldstein, J. L. Lebowitz, R. Tumulka, N. Zanghi, 
On the Distribution of the Wave Function for Systems in Thermal Equilibrium, 
quant-ph/0309021.

\bibitem{EI} A. Einstein and L. Infeld, {\it The evolution of Physics. From
early concepts to relativity and quanta,} Free Press, London (1967).

\bibitem{Bell} J. S. Bell, {\it Speakable and unspeakable in quantum mechanics,}
Cambridge Univ. Press (1987)

\bibitem{ESP} B. d'Espagnat, {\it Veiled Reality. An anlysis of present-day
 quantum mechanical concepts,} Addison-Wesley (1995).
 
\bibitem{Sh} A. Shimony, {\it Search for a naturalistic world view,} Cambridge Univ. Press (1993).

\bibitem{HS} H. P. Stapp, $S$-matrix interpretation of quantum theory, {\it Phys. Rev.} D
{\bf 3}, 1303-1320 (1971).

\end{thebibliography}
\end{document}